\documentclass[12pt,a4paper]{article}  
\usepackage{psfrag}   
\usepackage{graphics}  
\usepackage{amssymb,epsfig,amsmath,euscript,array} 
\makeatletter
\@addtoreset{equation}{section}
\makeatother

\newcounter{multieqs}

\newcommand{\be}{\begin{equation}}  
\newcommand{\ee}{\end{equation}}

\newcommand{\bm}[1]{\mbox{\boldmath $#1$}}

\def\bd{\begin{document}}  
\def\ed{\end{document}}  
\def\nn{\nonumber}  
\def\bea{\begin{eqnarray}}  
\def\eea{\end{eqnarray}}  
\let\bm=\bibitem  
\let\la=\label

\newcommand{\EQ}[1]{\begin{equation} #1 \end{equation}}  
\newcommand{\AL}[1]{\begin{subequations}\begin{align} #1 \end{align}\end{subequations}}  
\newcommand{\SP}[1]{\begin{equation}\begin{split} #1 \end{split}\end{equation}}  
\newcommand{\ALAT}[2]{\begin{subequations}\begin{alignat}{#1} #2 \end{alignat}\end{subequations}}  
\def\beqa{\begin{eqnarray}}   
\def\eeqa{\end{eqnarray}}   
\def\beq{\begin{equation}}   
\def\eeq{\end{equation}}   
 
\def\hf{{\textstyle{1\over2}}} 
\def\wbar{\bar w}
\def\mubar{\bar\mu}
\def\abar{\bar a}
\def\sigmabar{\bar\sigma}
\def\etabar{\bar\eta}
\def\zetabar{\bar\zeta}
\def\mubar{\bar\mu}
\def\nubar{\bar\nu}
\def\N{{\cal N}}  
\def\sst{\scriptscriptstyle}  
\def\thetabar{\bar\theta}  
\def\Tr{{\rm Tr}}  
\def\one{\mbox{1 \kern-.59em {\rm l}}}  
 \def\Nh{\hat{N}} 
  
\def\a{\alpha}      \def\da{{\dot\alpha}}  
\def\b{\beta}       \def\db{{\dot\beta}}  
\def\c{\gamma}  \def\G{\Gamma}  \def\cdt{\dot\gamma}  
\def\d{\delta}  \def\D{\Delta}  \def\ddt{\dot\delta}  
\def\e{\epsilon}        \def\vare{\varepsilon}  
\def\f{\phi}    \def\F{\Phi}    \def\vvf{\f}  
\def\h{\eta}  
\def\k{\kappa}  
\def\l{\lambda} \def\L{\Lambda}  
\def\m{\mu} \def\n{\nu}  
\def\o{\omega}  
\def\p{\pi} \def\P{\Pi}  
\def\r{\rho}  
\def\s{\sigma}  \def\S{\Sigma}  
\def\t{\tau}  
\def\th{\theta} \def\Th{\Theta} \def\vth{\vartheta}  
\def\X{\Xeta}  
\def\z{\zeta}  
  
\def\cA{{\cal A}} \def\cB{{\cal B}} \def\cC{{\cal C}}  
\def\cD{{\cal D}} \def\cE{{\cal E}} \def\cF{{\cal F}}  
\def\cG{{\cal G}} \def\cH{{\cal H}} \def\cI{{\cal I}}  
\def\cJ{{\cal J}} \def\cK{{\cal K}} \def\cL{{\cal L}}  
\def\cM{{\cal M}} \def\cN{{\cal N}} \def\cO{{\cal O}}  
\def\cP{{\cal P}} \def\cQ{{\cal Q}} \def\cR{{\cal R}}  
\def\cS{{\cal S}} \def\cT{{\cal T}} \def\cU{{\cal U}}  
\def\cV{{\cal V}} \def\cW{{\cal W}} \def\cX{{\cal X}}  
\def\cY{{\cal Y}} \def\cZ{{\cal Z}}

\def\ua{\underline{\alpha}}  
\def\ub{\underline{\phantom{\alpha}}\!\!\!\beta}  
\def\uc{\underline{\phantom{\alpha}}\!\!\!\gamma}  
\def\um{\underline{\mu}}  
\def\ud{\underline\delta}  
\def\ue{\underline\epsilon}  
\def\una{\underline a}\def\unA{\underline A}  
\def\unb{\underline b}\def\unB{\underline B}  
\def\unc{\underline c}\def\unC{\underline C}  
\def\und{\underline d}\def\unD{\underline D}  
\def\une{\underline e}\def\unE{\underline E}  
\def\unf{\underline{\phantom{e}}\!\!\!\! f}\def\unF{\underline F}  
\def\unm{\underline m}\def\unM{\underline M}  
\def\unn{\underline n}\def\unN{\underline N}  
\def\unp{\underline{\phantom{a}}\!\!\! p}\def\unP{\underline P}  
\def\unq{\underline{\phantom{a}}\!\!\! q}  
\def\unQ{\underline{\phantom{A}}\!\!\!\! Q}  
\def\unH{\underline{H}}  
  
\def\As {{A \hspace{-6.4pt} \slash}\;}  
\def\bs {{b \hspace{-6.4pt} \slash}\;}  
\def\Ds {{D \hspace{-6.4pt} \slash}\;}  
\def\ds {{\del \hspace{-6.4pt} \slash}\;}  
\def\ss {{\s \hspace{-6.4pt} \slash}\;}  
\def\ks {{ k \hspace{-6.4pt} \slash}\;}  
\def\ps {{p \hspace{-6.4pt} \slash}\;}  
\def\pas {{{p_1} \hspace{-6.4pt} \slash}\;}  
\def\pbs {{{p_2} \hspace{-6.4pt} \slash}\;}  
  
\def\Fh{\hat{F}}  
\def\Vh{\hat{V}}  
\def\Xh{\hat{X}}  
\def\ah{\hat{a}}  
\def\xh{\hat{x}}  
\def\yh{\hat{y}}  
\def\ph{\hat{p}}  
\def\xih{\hat{\xi}}  
  
\def\psit{\tilde{\psi}}  
\def\Psit{\tilde{\Psi}}  
\def\tht{\tilde{\th}}  
   
\def\At{\tilde{A}}  
\def\Qt{\tilde{Q}}  
\def\Rt{\tilde{R}}  
\def\Nt{\tilde{N}}  
  
\def\at{\tilde{a}}  
\def\st{\tilde{s}}  
\def\ft{\tilde{f}}  
\def\pt{\tilde{p}}  
\def\qt{\tilde{q}}  
\def\vt{\tilde{v}}  
\def\nt{\tilde{n}}  
  
\def\delb{\bar{\partial}}  
\def\bz{\bar{z}}  
\def\bD{\bar{D}}  
\def\bB{\bar{B}}  
  
\def\bk{{\bf k}}  
\def\bl{{\bf l}}  
\def\bp{{\bf p}}  
\def\bq{{\bf q}}  
\def\br{{\bf r}}  
\def\bx{{\bf x}}  
\def\by{{\bf y}}  
\def\bR{{\bf R}}  
\def\bV{{\bf V}}  
  
\def\d{\delta}\def\D{\Delta}\def\ddt{\dot\delta}  
  
\def\pa{\partial} \def\del{\partial}  
\def\xx{\times}  
\def\uno{\mbox{1 \kern-.59em {\rm l}}}    
  
\def\trp{^{\top}}  
\def\inv{^{-1}}  
\def\dag{{^{\dagger}}}  
\def\pr{^{\prime}}  
  
\def\rar{\rightarrow}  
\def\lar{\leftarrow}  
\def\lrar{\leftrightarrow}  
  
\newcommand{\0}{\,\!}      %this is just NOTHING!  
\def\one{1\!\!1\,\,}  
\def\im{\imath}  
\def\jm{\jmath}  
  
\newcommand{\tr}{\mbox{tr}}  
\newcommand{\slsh}[1]{/ \!\!\!\! #1}  
  
\def\vac{|0\rangle}  
\def\lvac{\langle 0|}  
  
\def\hlf{\frac{1}{2}}  
\def\ove#1{\frac{1}{#1}}  

\def\Box{\square}  
\def\ZZ{\mathbb{Z}}  
\def\CC#1{({\bf #1})}  
\def\bcomment#1{}  
\def\bfhat#1{{\bf \hat{#1}}}  
\def\VEV#1{\left\langle #1\right\rangle}  

\newcommand{\ex}[1]{{\rm e}^{#1}} \def\ii{{\rm i}}  

\def\rr{{\rm r}} \def\rs{{\rm s}}\def\rv{{\rm v}}
\def\ri{{\rm i}}\def\rj{{\rm j}}
  
\newcommand{\lrbrk}[1]{\left(#1\right)}
\newcommand{\sfrac}[2]{{\textstyle\frac{#1}{#2}}}

\font\mybb=msbm10 at 12pt
\def\bb#1{\hbox{\mybb#1}}

\font\myBB=msbm10 at 18pt
\def\BB#1{\hbox{\myBB#1}}

\begin{document}  
  
\hfill{ CERN-TH/2003-273} 
   
\hfill{ hep-th/0311065}  
   
\vspace{20pt}   
   
\begin{center}  
  
{\Large \bf From Branes to Branes}

\vspace{30pt}

{\bf Valentin V.~Khoze}

\medskip

{\small \em
Theory Division,
CERN,\\
CH-1211 Geneva 23, Switzerland 
}

{\small  and}

{\small \em
Centre for Particle Theory,
Department of Physics and IPPP,\\
University of Durham, Durham, DH1 3LE, UK
}

\vspace{10pt}  
  
{\sffamily \tt valya.khoze@durham.ac.uk }

\vspace{30pt}  

{\bf Abstract}

\end{center}  
We use the `branes within branes' approach 
to study the appearance of stable $(p-2)$-branes and unstable $(p-1)$-branes
in type II string theory from $p$-brane--$p$-antibrane pairs.
Our goal is to describe the emergence of these lower dimensional branes
from brane-antibrane pairs in string theory using a tractable gauge theory language. 
This is achieved by suspending the original $p$-brane--$p$-antibrane pair 
between two $(p+2)$-branes, and describing its dynamics in terms of the 
worldvolume gauge theory on the spectator $(p+2)$-branes. Instantons,
monopoles, sphalerons and their higher-dimensional generalizations
in this worldvolume gauge theory correspond to stable (BPS) and unstable 
(non-BPS) branes in string theory. Collisions of stable branes with corresponding antibranes
and production of lower-dimensional branes in string theory are described 
in a straightforward way in gauge theory. 
Tachyonic modes on the $p$-brane--$p$-antibrane
worldvolume do not appear in our analysis since we work on the worldvolume
of the spectator $(p+2)$-branes. Our results on brane descent relations
are in agreement with Sen's tachyon condensation approach.

\vspace{0.5cm}  
  
\setcounter{page}{0}  
\thispagestyle{empty}  
\newpage

%%%%%%%%%%%%%% ordinary document (end) %%%%%%%%%%%%%%%%%%%%%%%%%%%%%%%%
%\parindent=0pt

\section{Introduction}

Much of our current quantitative 
understanding of non-perturbative string theory centers on stable BPS-saturated
D-branes and appeals to powerful 
constraints imposed by supersymmetry. One way to go beyond this and to
introduce inter-brane interactions and time-dependent processes involving branes
is to consider non-supersymmetric  brane-antibrane systems. 
D-branes and D-antibranes carry opposite RR charges and are not protected
by supersymmetry. These branes and antibranes are expected to scatter and 
to annihilate each other in string theory
as particles or extended objects do in quantum field theory.

Most of the recent progress in this subject follows the
approach initiated by Sen \cite{Sen1,Sen,Sen-d,Sen3} and based on tachyon condensation
in brane-antibrane systems. In this approach the perturbative instability of the
brane-antibrane pair at short distances \cite{BS} manifests itself as the 
appearance of a tachyonic  mode of the fundamental F1
string which is stretched
between the brane and the antibrane. This gives rise to a tachyon
field, which is a complex scalar
living on the worldvolume of the coincident brane-antibrane configuration,
and
transforming in the bi-fundamental $\bar{U}(1) \times U(1)$ representation
of the gauge theories on the brane and the antibrane.
The idea is that the tachyon condenses 
and causes the annihilation of the brane-antibrane
configuration to the vacuum. 
More precisely, Sen conjectures that the tachyon potential is of the
'Mexican Hat' type such that
the unstable configuration at the top of the
potential corresponds to the coincident brane-antibrane configuration,
and the ground state is the closed string vacuum
with no branes or open strings left. 

Importantly, stable and unstable
lower-dimensional D-branes 
can now appear as solitons in the tachyon field
\cite{Sen}. Stable branes appear as
co-dimension-2 topological solitons in the brane-antibrane worldvolume, and the
unstable branes are co-dimension-1 unstable classical solutions.
This implies that all branes in e.g. type II string theory can be obtained 
from annihilations of the highest-dimensional 
D9 and anti-D9-branes. An elegant brane classification
follows from this \cite{Witten-k} and is based on K-theory.

In Sen's scenario, the complex tachyon serves as 
the Higgs field which spontaneously breaks the $U(1) \times U(1)$ gauge theory
on the brane-antibrane worldvolume to a diagonal $U(1)$. The fate of this remaining 
$U(1)$ is a little less clear, as it is supposed to have completely disappeared
in the string vacuum. It was argued in \cite{Yietal} that the diagonal $U(1)$ is
not seen in the vacuum because it is
confined due to the condensation of different tachyons living this time on the
D-branes stretching between the brane-antibrane pair.

The main motivation of this paper is to gain 
further insights into brane-antibrane systems and
to find independent confirmations
of the 
results of Sen's approach
using a different language --- gauge theory.

Very recently the authors of \cite{HTaylor} have made
interesting progress in this direction by seeing signs of
the tachyon condensation from the gauge theory perspective.
More precisely 
Ref. \cite{HTaylor}
studied the gauge theory living on the worldvolume of two 
intersecting branes. When the intersection angle $\theta$ is close to $\pi$,
the configuration becomes the brane-antibrane pair. The difficulty in  
using the results of this
approach is that for $\theta \sim \pi$ the worldvolume theory is not described by
a gauge theory. Hence, the authors of \cite{HTaylor} had
to work with small values of $\theta$ (which is more like a brane-brane
configuration rather than a brane-antibrane)
where the gauge theory description
is valid, and to extrapolate their findings to a regime of interest, 
$\theta \sim \pi$.
We will avoid this difficulty altogether
by introducing additional -- spectator -- branes
and working with the gauge theory on their worldvolume.

We will use a version of a `branes within branes' approach 
where the $p$-brane--$p$-antibrane pair is suspended between 
two $(p+2)$-branes. The dynamics of the $p$-brane--$p$-antibrane annihilation
can then be described in terms of the worldvolume gauge theory 
on the $(p+2)$-branes. 
Descent relations between stable branes which follow from Sen's approach
\cite{Witten-k,Sen-d} imply that lower-dimensional branes can be produced 
in brane-antibrane annihilations. We
want to understand this in a gauge theory language, where 
one might naively expect that, for example,
instanton-antiinstanton configurations and monopole-antimonopole configurations
annihilate each other completely into a perturbative vacuum. 

In fact, Taubes \cite{Taubes-lect,Taubes} showed long time ago that monopole-antimonopole 
classical configurations can be used to construct non-contractible loops which
(as will be explained below) give rise to instanton and sphaleron solutions.
This observation of Taubes will be at the heart of the SYM analysis in this paper.

In Section 2 we will outline the Taubes construction 
applied to the $\cN=4$ SYM and
explain how it
links together all three types
of the brane solutions in SYM in four dimensions. 
In the second half of Section 2 we 
switch to type IIB string theory and embed gauge instantons, monopoles and sphalerons
into string theory
as D-branes within branes. This allows us to embed the non-contractible
monopole-antimonopole loop in string theory.
 
In the second half of the paper (Section 3) we will explain
how to generalize and incorporate this construction 
to higher-dimensional D-branes in type II string theory.
We will describe how the lower-dimensional stable branes are
produced in brane-antibrane annihilation processes and derive
 the descent relations between branes.
We will also explain how these relations incorporate stable non-Dirichlet
branes, such as F1, NS5 and the S-dual of the D-instanton.
Section 4 presents our conclusions and some open questions.

\section{Instantons and sphalerons from monopoles in SYM and in string theory}

The field theory considered in this section
is the $\cN=4$ supersymmetric $SU(2)$ gauge theory
in Minkowski and also in Euclidean 4-dimensional spacetimes. Greek indices, $\mu,\nu$, will
refer to spacetime components, $\mu,\nu=0,1,2,3$ in Minkowski and $\mu,\nu=1,2,3,4$
in Euclid. Latin indices, $m,n$, label spatial directions, $m,n=1,2,3$.
This gauge theory in the Coulomb phase
is embedded in type IIB string theory 
at low energies ($\alpha'\rightarrow 0$) as the worldvolume 
theory on two parallel D3-branes with the relative separation $2\pi \alpha' v$
along the perpendicular to the branes direction, e.g. $x_9$.

\subsection{The monopole}

First we recall some basic facts about 't Hooft--Polyakov monopoles
\cite{Thooft-m,Polyakov,BPS}. 
The standard BPS monopole solution in a static Hedgehog gauge is \cite{BPS}
\begin{equation}\begin{split}
\Phi^{\sst\rm mono} (x_n)\ &= \ {1\over g}\,\big(gv|x| \ {\rm coth}(gv|x|) -1
\big) 
{x_a \over |x|^2}\,{\tau^a \over 2} \ , \\
A^{\sst\rm mono}_m (x_n)
 \ &= \ {1\over g}\,\Big(1-{gv|x| \over {\rm sinh}(gv|x|)}
\Big)\epsilon_{mn a}{x_n \over |x|^2}\,{\tau^a \over 2}
 \ .
\label{bpsmo}\end{split}\end{equation}
Here $v$ is the vacuum expectation value (vev) of the adjoint scalar field $\Phi$,
which follows from the large-$|x|$ asymptotics of the solution, 
$\Phi^{{\sst\rm mono}\,(a)} \rightarrow v x^a/|x|$,
the distance in 3D space is denoted as 
$|x|=\sqrt{x_m x_m}$, and $\tau^a$ are the three Pauli matrices. 

The configuration \eqref{bpsmo} can be embedded into the $\cN=4$ SYM theory as
\be 
\phi_1 = \Phi^{\sst\rm mono}(x_n) \quad , \quad 
\phi_{2}=\phi_{3}=\ldots=\phi_{6} = 0 \quad , \quad
A_0 =0 \quad , \quad A_m=A^{\sst\rm mono}_m(x_n) \ .
\label{n4mono}
\ee
Here the $\cN=4$ SYM is on the Coulomb branch with one of the six
real scalar fields, i.e. $\phi_1$, having a nonzero 
vev\footnote{R symmetry
$SO(6)$-rotations can always be used to single out any one of the six
scalars. For definiteness we will always choose to give the vev to $\phi_1$.}
$v$. This breaks gauge $SU(2)$ spontaneously to $U(1)$.
Expressions \eqref{n4mono} 
give a time-independent  classical solution,
with finite energy,
and one unit of magnetic charge of the unbroken $U(1)$.
The monopole solution \eqref{bpsmo} or \eqref{n4mono} is topological in nature, 
its magnetic charge
is the winding number of $S^2 \rightarrow S^2$. Here the first $S^2$ is the 2-sphere
at the boundary of $R^3$ as $|x| \rightarrow \infty,$ and the second
$S^2 = SU(2)/U(1)$ is the manifold of the asymptotic values 
of the scalar field 
$\phi_1^a$ as $\sum_{a=1}^3 \langle(\phi_1^a)^2\rangle =v^2$.

As already mentioned, the monopole solution \eqref{bpsmo} is written
in the static Hedgehog gauge,
\be
\langle A_0\rangle =\, 0 \qquad , \qquad
\langle \phi_1^a\rangle =\, v \,x^a/|x| \ .
\label{hedge}
\ee
It is more convenient for our purposes
to gauge-transform \eqref{bpsmo} to the unitary gauge 
where
\be
\langle \phi_1^a \rangle =\, v\, \delta^{a3} \ .
\label{unit}
\ee
This can be achieved by using a gauge transformation which  
transforms the unit vector $x^a/|x|$ into the
unit vector along the third direction, $\delta^{a3}$, \cite{Thooft-m}.
This gauge-transformation is singular along the ray emitted from the
monopole centre and  introduces the Dirac string into the regular monopole configuration
\eqref{bpsmo}. This string,
however, is clearly a gauge artifact and will be largely ignored
in what follows.

The scalar field of the monopole in the singular unitary gauge is purely
Abelian, i.e. aligned with the vev in \eqref{unit}, and is of the form
\be
\phi_1 = \Phi^{\sst\rm mono}(x_n)=
 \ {1\over g}\,\big(g v|x| \ {\rm coth}(g v|x|) -1
\big) 
{1 \over |x|}\,{\tau^3 \over 2} \ ,
\label{scmonoun}
\ee
while the gauge field $A_m$ contains Abelian $(\propto \tau^{3})$
as well as non-Abelian
$(\propto \tau^{1,2})$ components.
The unitary gauge form of the monopole \eqref{scmonoun} will  be
required  below for a realization of the monopole in type IIB string theory
as a D1-brane suspended
between the two parallel D3-branes. 

For future reference we note here that
the SYM-monopole is a BPS-saturated configurations
in the sense that the monopole is
annihilated by half of the sixteen supercharges of the $\cN=4$ theory.
An easy way to see it is to note that if one renames the scalar monopole
component in \eqref{bpsmo} as $A_4$, the corresponding field-strength 
$F_{\mu\nu}$ (made out of $A_n,A_4$) is self-dual,
$F_{\mu\nu}= {}^*F_{\mu\nu}$ for the monopole solution. This implies that
eight of the supercharges of the $\cN=4$ theory will annihilate the bosonic
monopole, and further eight supercharges will give eight adjoint fermion
zero modes of the monopole. There are precisely two fermion zero modes
for each of the four flavours $I=1,\ldots,4$:
\begin{equation}\lambda^{I}_\alpha \ = \ \hf \xi^I_\beta
(\sigma^\mu \sigmabar^\nu)_\alpha^{\ \beta} 
F_{\mu\nu} \ , \label{lss}\end{equation}
where $\xi^I_\beta$ are the Grassmann collective coordinates for the four
spinor supercharges $Q_I^\beta$ and
$\sigma^\mu$ and $\sigmabar^\nu$ are the four Pauli matrices.
 
So far we have been discussing the monopole with magnetic charge $+1$.
Being BPS-saturated,
single monopoles do not interact with each other and multi-monopole
configurations can be constructed.
General multi-monopole solutions follow from the Nahm construction \cite{Nahm}.
There are also antimonopole solutions with negative magnetic charges,
they are obtained from monopole solutions by switching the sign of
the scalar field. We also note that there is a Coulomb attraction between
monopoles and antimonopoles at large distances, 
and no static monopole-antimonopole configuration
exists as a classical solution. There are, of course, time-dependent 
monopole-antimonopole solutions which describe a classical scattering
process. Such solutions, in principle, can be constructed numerically.

\subsection{The instanton}

The single-instanton configuration \cite{BPST,Thooft-i} in Euclidean $\cN=4$ theory is 
\begin{equation}\begin{split}
A^{\sst\rm inst}_\mu (x)
 \ &= \ {2\over g}\,{\rho^2 \over x^2(x^2+\rho^2)}\, \bar{\eta}^a_{\mu\nu}\,
 x_\nu \, {\tau^a\over 2}
  \ , \\
\phi_1^{\sst\rm inst} (x)\ &= \ v\,{x^2 \over x^2+\rho^2}\, {\tau^3\over 2} \ ,
\label{inst}\end{split}\end{equation}
where $\bar{\eta}^a_{\mu\nu}$ is the 't Hooft $\bar{\eta}$-symbol \cite{Thooft-i},
and $\rho$ is the instanton scale size.\footnote{Other bosonic collective coordinates
of the $SU(2)$ instanton are the global $SU(2)$ rotations of the gauge field only,
and the four-translations of the gauge and the Higgs field together. Together with
$\rho$ this makes $1+3+4=8$ bosonic zero modes.}
In \eqref{inst} we showed only the non-vanishing bosonic fields and
set fermionic collective coordinates to zero. 
For future reference, we note that the instanton scalar field in \eqref{inst} 
is purely Abelian and is already in the unitary gauge \eqref{unit}.

The instanton \eqref{inst} is a time-dependent field configuration
with finite Euclidean action,
\be
S_E = {8\pi^2 \over g^2} + 4 \pi^2 v^2 \rho^2 \ .
\ee
For a non-zero vev,  $S_E $ explicitly depends on $\rho$ and thus,
$\rho$ is not an exact zero mode of the instanton, and 
\eqref{inst} is not an exact solution of equations of motion 
for $\rho>0$ (at $\rho=0$ the instanton is singular). We recall that
for nonzero vev  a nontrivial regular solution cannot exist, due to
Derrick's theorem: for any putative solution one can lower the action further
simply by shrinking the configuration. One way to fix this
problem was found by Affleck \cite{Affleck}.
For a brief practical review with an application see sections 3 and 4 of 
\cite{DKM-mo1}. The idea is as follows: a
new operator, or Affleck constraint, is introduced 
into the action
by means of a Faddeev-Popov insertion of unity. 
If this operator is of suitably high dimension,
Derrick's theorem is avoided, and the instanton stabilizes at a fixed scale size $\rho.$
The integration over the 
Faddeev-Popov Lagrange multiplier in the path integral
can then be traded off for the
integration over $\rho.$ 
The now-stable solutions are known as constrained instantons.

The detailed 
shape of the constrained instanton depends on a choice of constraint. 
But it turns out that only the short-distance
regime, $x\ll1/M_W,$
and the long-distance regime, $x\gg1/M_W,$
of the instanton are important.
(Here $M_W=gv$ is the $W$-boson mass.) In particular, the instanton
measure and action depend only on the short-distance instanton \eqref{inst},
while the low-energy fields in the instanton background 
require the long-distance instanton --
see \cite{DHKM-rept} and references therein.

As in the monopole case earlier,
the instanton solution is topologically stable, but instantons are governed
by $S^3$ spheres. Instanton topological charge,
\be
Q={1\over 16\pi^2}\int F_{\mu\nu}  {}^* F_{\mu\nu} \, d^4x \ ,
\label{Qinst}
\ee
is the winding of $S^3 \rightarrow S^3$
also known as Pontryagin number. The first $S^3$ is the large-$|x|$ sphere of 
Euclidean 4D spacetime, and the second $S^3$ is $SU(2)$ (it arises from the requirement
that $A_\mu$ goes to a pure gauge at large values of $x$ as a necessary condition
for the finiteness of the action).

Since the instanton field-strength is self-dual, instantons are BPS-saturated.
As in the monopole case, there are precisely eight exact (adjoint) fermion zero modes
in the instanton background.\footnote{For the single instanton there are also 
eight quasi-zero fermion modes,
which are lifted by the vev $v$.}
Multi-instanton solutions follow from the ADHM formalism 
\cite{ADHM,CGFT,CWS} -- see
\cite{DHKM-rept} for a review and applications.
Finally, instanton-antiinstanton configurations are not classical solutions
at finite separations; instantons and antiinstantons interact and annihilate
into a perturbative vacuum. This is different from the
time-dependent monopole-antimonopole case 
earlier.\footnote{Instanton-antiinstanton configurations at finite separations
can be seen and rigorously defined as solutions to 
 the valley equation of Yung -- see \cite{Y,KR} for the formalism and applications
 to gauge theories.}
 
\subsection{Barrier penetration and sphalerons}

In gauge theory, the instanton solution in the $A_4=0$ gauge mediates the transition
between two topologically distinct vacua -- e.g. from a trivial vacuum to a vacuum 
with a winding number one. Instanton contributions to Euclidean path integrals 
correspond to tunneling transitions between these two vacua \cite{JRCDG}.

When the vev is non-zero, there is a barrier between the vacua which corresponds to
an unstable classical solution, the sphaleron \cite{Manton}.
The sphaleron solution can be determined as the maximal energy
configuration along the non-contractible loop of field configurations starting and
ending in the vacuum. More concretely, consider a continuous path made of finite
energy field configurations, which starts at the trivial vacuum and terminates in
the vacuum with winding number one. Find the point with maximal energy on each of 
these paths and find such a path where this energy is minimal. This `minimax' 
procedure determines the saddle-point solution on top of the minimal energy path.
This solution is a sphaleron and it has precisely one negative mode.
When the two vacua at the beginning and at the end of the path are identified,
the path becomes a non-contractible loop as depicted in Figure 1.
\begin{figure} [ht]
\label{fig1}
\begin{center}
{\scalebox{0.5}{
\includegraphics{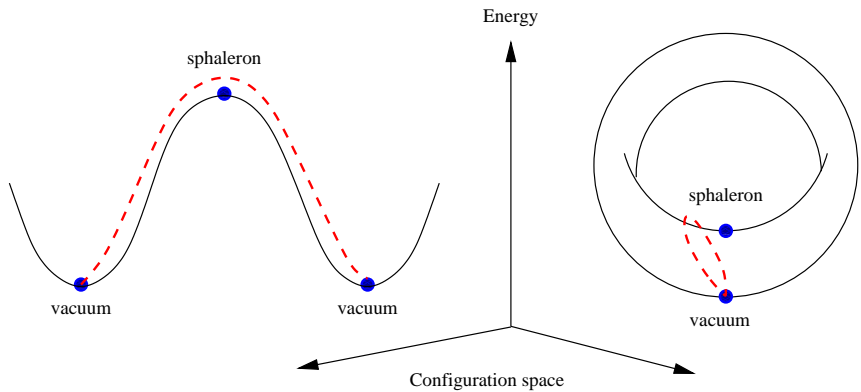}}
}
\end{center}
\caption{\small \it A path made of finite-energy classical field
configurations
interpolating between two topologically distinct vacua.
When the vacua are identified
in the picture on the right, the path becomes a non-contractible loop. 
The loop with the minimal value of the maximal
energy will pass through the sphaleron solution. 
}
\end{figure}

This discussion can be complemented by an argument due to 
Taubes \cite{Taubes} who has shown 
rigorously that the sphaleron solution exists 
in the gauge theory with an adjoint
Higgs. This establishes the existence of the sphaleron solution in the
$\cN=4$ SYM theory. 
An explicit form of the sphaleron in $\cN=4$
can in principle be found numerically by
choosing a suitable family of non-contractible loops.

It is clear from the above discussion that the sphaleron is not
a topological solution, it is unstable as it decays 
along the non-contractible loop to the vacuum.
At the same time, the non-contractible loop itself has a topology
of the instanton. Hence, instantons and sphalerons are intimately related
to each other via the notion of non-contractible loops. The instanton corresponds
to the loop with the minimal Euclidean action, and the sphaleron to
the loop with the minimal maximal energy. 
What is most remarkable, however, is the observation of Taubes 
\cite{Taubes-lect,Taubes} that the non-contractible loop itself
is constructed from the monopole-antimonopole pair. 

In the following subsection we will outline the Taubes construction and
explain how it
links together all three types
of the brane solutions in the $\cN=4$ SYM.

\subsection{Non-contractible monopole loop in gauge theory}

We will now describe the key idea which is
at the heart of the Taubes construction  
\cite{Taubes-lect,Taubes} of a non-contractible monopole loop in a gauge theory with
an adjoint scalar field. Applied directly to the $\cN=4$ SYM in the Coulomb
phase, this construction derives the very existence of  sphaleron and instanton
solutions in this theory starting from the monopole solution (or more precisely,
a combination of 't Hooft-Polyakov monopoles 
with the net magnetic charge equal to zero). 
In Section 3
this line of reasoning will become our starting point in
explaining how lower-dimensional branes are produced in annihilations of  
brane-antibrane pairs of higher dimension.

The idea of Taubes was to construct a representative of a homotopy class of non-contractible
loops in the $\cN=4$ classical field configuration space
from the 't Hooft-Polyakov monopole solution \eqref{bpsmo}. 
The element of this homotopy class which has the lowest Euclidean
action along the loop is the instanton solution, and the element
with the lowest maximal energy of a point on the loop is the sphaleron saddle-point
solution.

First we consider the Hedgehog gauge where the
topology is more obvious, and then will recast the same argument in the unitary
gauge which is more
suited for branes within branes applications in string theory.

We recall from \eqref{bpsmo} that at large distances, $x \gg 1/M_W,$ from the monopole
centre, the monopole scalar field 
is an $S^2$ hedgehog while the antimonopole has the reversed picture
of field-lines, see Figure 2. 
\begin{figure} [ht]
\label{fig2}
\begin{center}
{\scalebox{0.4}{
\includegraphics{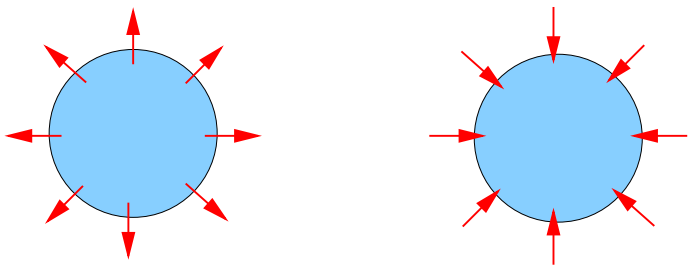}}
}
\end{center}
\caption{\small \it Monopole and antimonopole scalar field components in the hedgehog gauge.}
\end{figure}
Now consider a composite configuration made out of the monopole and the antimonopole
at a large separation as shown on Figure 3.  This configuration is a 
fixed-time snap-shot of the corresponding time-dependent classical solution with the
net monopole charge zero. 
If brought together, the monopole field would cancel 
the antimonopole precisely, leaving perturbative vacuum (plus radiation if the
collision occurs in real time). Now we want to continuously  deform the configuration
in a non-trivial way. We start rotating the monopole along the axis of the 
configuration throat 
where the field-lines match, while keeping the antimonopole fixed. At any time during 
this rotation, the long-range (Abelian) fields still match, but not the short-range
(non-Abelian) fields. Thus, the monopoles would not annihilate if brought together
since their non-Abelian fields would not cancel. This remains so until the rotation
completes a full circle, and we arrive
at the original configuration where all fields match and which can be shrunk
to a perturbative vacuum. 
\begin{figure} [ht]
\label{fig3}
\begin{center}
{\scalebox{0.4}{
\includegraphics{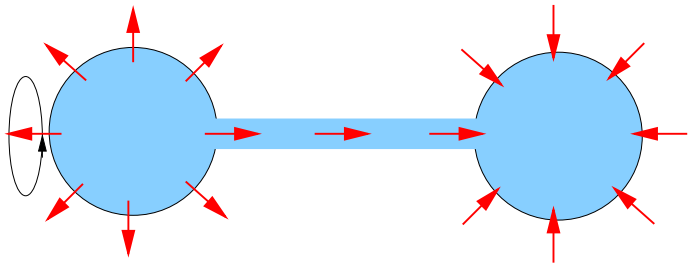}}
}
\end{center}
\caption{\small \it Non-contractible monopole-antimonopole
loop in the hedgehog gauge. The monopole is rotated by $2\pi$ while the
antimonopole is fixed. }
\end{figure}
By starting in the vacuum, then continuously deforming it
to produce a classical monopole-antimonopole pair, then separating the pair,
then rotating one of
the monopoles by $2\pi$ and, finally, bringing them together to annihilate, we
create a non-contractible monopole-antimonopole loop in the classical configuration space.
Since the antimonople can be described as the monopole moving backwards in time,
this loop can also be viewed as a single monopole making a full circle 
in space, rotating at the same time by a full rotation.\footnote{In view of this
equivalence we will use the terms `non-contractible monopole loop' and 
`non-contractible monopole-antimonopole loop' interchangeably.}

Remarkably, it turns out that this non-contractible loop has the topology of the 
instanton.
The $S^3$-sphere associated with the instanton is a twisted product of $S^1$ 
(the monopole field rotation) and the $S^2$-sphere formed by the monopole scalar 
hedgehog field. $S^3$ exhibited as a twisted $S^1$ bundle over $S^2$ is the
Hopf fibration. The topological charge associated with the non-contractible monopole
loop is the Pontryagin index \cite{Taubes-lect}.

The same picture can also be described in physical terms in the unitary gauge. 
In the isospin space the monopole in the unitary
gauge has Abelian (isospin-3) and non-Abelian (isospin-1,2) field components
as indicated on Figure 4.
The monopole and the antimonopole have long-range Abelian and short-range
non-Abelian interactions. The isospin-3 interaction, 
$V_{\rm long}(r)= -2/r,$ is long-range
and is always attractive.\footnote{The factor of two comes 
from adding the Higgs-mediated  to the gauge-mediated Coulomb interaction.}
The isospin-1,2 interaction is short-range, but also
depends on the relative 
orientation $\theta$ of the isospin-1 and isospin-2 components. Attraction
changes to repulsion as $\theta$ varies from $-\pi$ to $0$.
\begin{figure} [ht]
\label{fig4}
\psfrag {theta} {\LARGE ${\theta}$}
\psfrag {A1} {\LARGE${A_\mu^{(1)}}$}
\psfrag {A2} {\LARGE${A_\mu^{(2)}}$}
\psfrag {A3} {\LARGE${A_\mu^{(3)}\, ,\,\phi^{(3)}}$}
\psfrag {M} {\LARGE Monopole}
\psfrag {AM} {\LARGE Antimonopole}
\begin{center}
{\scalebox{0.5}{
\includegraphics{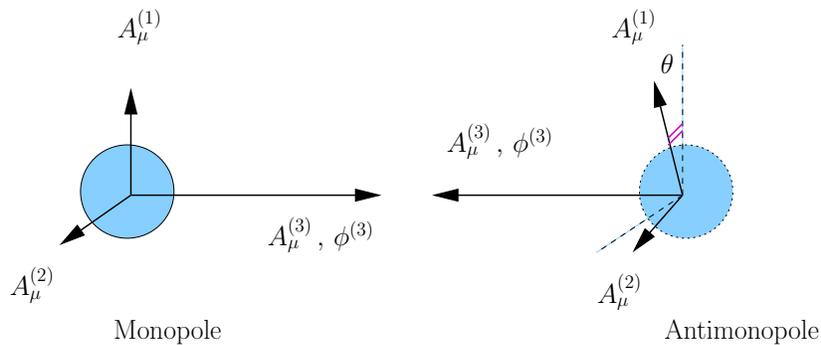}}
}
\end{center}
\caption{\small \it Monopole and antimonopole field components in the unitary gauge
}
\end{figure}
The total potential energy at intermediate distances can be represented
as follows \cite{Taubes-lect}
\be 
V(r)= -{2 \over r} \left(1-e^{-M_W r}\left(\hf +\cos \theta\right)\right) \ .
\label{itsmm}
\ee
The non-contractible loop, 
\be
l(\tau) =(A_\mu^\tau(x_n), \phi^\tau(x_n)) \ ,
\ee
is a continuous family of static finite-energy configurations
parameterized by $\tau$, see Figure 5.
At the initial value of $\tau=\tau_0$ one starts 
from the vacuum, $(0,v {\tau^3\over 2}),$ and as $\tau$ increases, 
creates from the vacuum the monopole-antimonopole configuration. 
The monopole-antimonopole parametrization 
$(r,\theta)$ is initially $(0,-\pi)$ and changes to $(R,-\pi)$
as $\tau$ grows from $\tau_0$ to $\tau_1$. As $\tau$ continues to
increase 
we keep $r=R$ fixed and gauge-rotate
the monopole by increasing $\theta$ continuously from $-\pi$ to $0$
at $\tau=\tau_2$, and then further to $+\pi$ 
at $\tau=\tau_3$.
Finally, after completing the full rotation of the monopole, we bring the
configuration $(R,\pi)$ to the vacuum $(0,\pi)$ as $\tau$ reaches its
final value $\tau_4$.
This loop is non-contractible since for any fixed value of $\tau$ in the
vicinity of $\tau_2$, the monopoles cannot be brought together, as their 
non-Abelian interaction is repulsive.
Of course this loop is completely identical to the loop in the Hedgehog gauge 
discussed earlier.
\begin{figure} [ht]
\label{fig5}
\psfrag {T1} {\LARGE${\tau_0 \rightarrow \tau_1}$}
\psfrag {T2} {\LARGE${\tau_1 \rightarrow \tau_3}$}
\psfrag {T3} {\LARGE${\tau_3 \rightarrow \tau_4}$}
\begin{center}
{\scalebox{0.5}{
\includegraphics{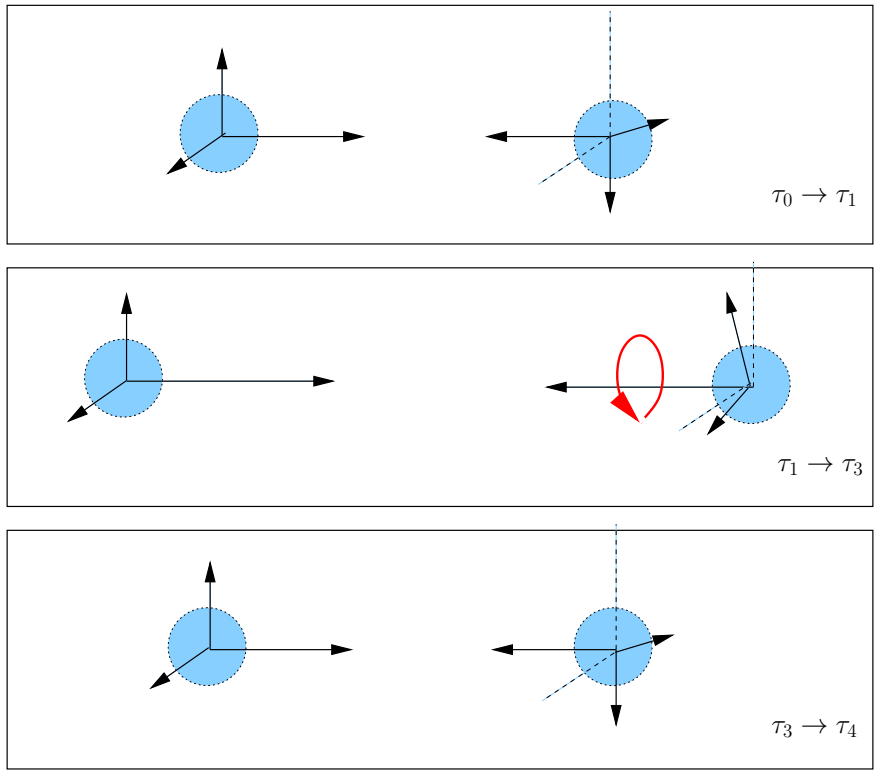}}
}
\end{center}
\caption{\small \it Monopole-antimonopole non-contractible loop in the unitary gauge.
}
\end{figure}
When the loop parameter $\tau$ is identified with the Euclidean time $x_4$,
the non-contractible loop is in the same homotopy class as the 1-instanton
solution; its topology is characterized by Pontryagin number equal to one.
When $\tau$ is interpreted as $x_4$ one can calculate the Euclidean action, $S_E,$
along the loop as $\int d\tau E$. Since the energy $E$ of the configuration
for each $\tau$ is finite, and since $\tau$ varies in a finite interval, 
the action $S_E$ is finite for our loop. Now, by continuously deforming the 
trial loop described above,
one can find the loop which minimizes $S_E$. It is the instanton.

Similarly, in the same homotopy class we can look for the loop 
which now has the lowest maximal energy,
i.e. for every loop find the value of $\tau$,
$\tau=\tau_*$, for which the energy,$E$ is maximal
(for our trial loop it is at $\tau_*=\tau_2$)
and then choose the loop which minimizes the value of $E(\tau_*)$.
The corresponding
configuration is a static saddle-point solution of equations of motion.
It is the sphaleron, and its energy represents the top of the barrier
under which the instanton tunnels.
 
In conclusion we comment that in this construction, the loop parameter
$\tau$ is not to be interpreted as the real time $x_0$. The real-time
classical process of monopole-antimonopole creation and subsequent annihilation
cannot start from the vacuum as the energy is not conserved. In fact,
the energy varies continuously along the loop, from zero to the sphaleron
mass and then back to zero. This also fits with the fact that the instanton
is the imaginary-time solution of classical equations.
We will add the real time dimension to this discussion in Section 3.

\subsection{Monopoles and instantons as branes within branes}

Some of the most remarkable developments in instanton calculus in gauge theory
came with the realization that the 
ADHM formalism \cite{ADHM,CGFT,CWS} arises naturally in the context of string theory
\cite{Witten-i1,Witten-i2,Douglas-i1,Douglas-i2,DHKMV-mo3}.  
Instantons
in the $\cN=4$ gauge theory in four dimensions correspond precisely to the boundstates
of D$(-1)$-branes on the worldvolume of coincident or parallel D3-branes. 
Instanton solutions in gauge theory are localized objects in space and time,
and so is the point-like D$(-1)$-brane within the 4-dimensional worldvolume of 
D3-branes.
More generally, the standard 4D instanton solutions embedded in a higher-dimensional
gauge theory, are realized in string theory as D$p$-branes within D$(p+4)$-branes.
The low energy collective
dynamics of $N$ coincident D$(p+4)$-branes in Type II string theory
is described by a $U(N)$ SUSY gauge theory in $p+5$-dimensions with 16
supercharges. An instanton in the worldvolume
theory of the D$(p+4)$-branes is a soliton which
has 4 transverse directions in the higher dimensional brane, i.e. it is a $p$-brane.
Remarkably, it is precisely a D$p$-brane bound to the
D$(p+4)$-branes. In general $k$ D$p$-branes
bound to the $N$ higher dimensional D$(p+4)$-branes correspond to a charge
$k$ instanton in a $U(N)$ SUSY gauge theory. The gauge theory and the D-brane
realizations of instantons are both BPS-saturated configurations.

Not only the ADHM multi-instanton
gauge field can be re-derived in string theory using a brane-probe approach
\cite{Witten-i2,Douglas-i2}, but also the $k$-instanton integration measure and action
in the $U(N)$ $\cN=4$ gauge theory
is identical to the partition function of $k$ D$(-1)$-branes within the 
$N$ D3-branes in type IIB string theory \cite{DHKMV-mo3} (for a review see
Section 10 of \cite{DHKM-rept}). 

For now let us set $p=-1$ so that the worldvolume theory on D$(p+4)$-branes
is a (3+1)-dimensional gauge theory.
For two coincident D3-branes we have the $U(2)$ (or decoupling the 
overall $U(1)$ factor, the $SU(2)$) $\cN=4$ gauge theory in the conformal 
phase. The instanton is the D$(-1)$-brane lying within the worldvolume
of the coincident D3-branes. An interesting question to ask is what
happens to this geometrical picture when the two D3-branes are separated,
i.e. when we the $\cN=4$ SYM develops a nonzero vev $v$. In other words,
for a general $p$, how is the $(p+1)$-dimensional worldvolume of the D$p$-brane
situated in relation to the separated 
$(p+5)$-dimensional worldvolumes of two D$(p+4)$-branes?

Before addressing this it will be useful
to recall the
realization of the monopole in type IIB string theory.
The monopole is a D1-brane suspended
between the two parallel D3-branes. This is a BPS configuration
in string theory as it preserves eight supersymmetries, just as
the $\cN=4$ monopole.
The worldvolume theory on D3-branes
is the $\cN=4$ gauge theory, 
and the two diagonal elements of
$2\pi \alpha' \langle \phi_1 \rangle = 2\pi\alpha' v \,\tau^3/2 $
are identified 
with the positions of the D3-branes along the 
external direction. On the D3-brane worldvolume the ends of the D1-brane
span the world-line of a particle -- the SYM monopole.

Similarly to the instanton case, this realization
of the monopole is straightforwardly generalized
by T-duality to a D$p$-brane stretched between two D$(p+2)$-branes.

The precise correspondence between the boundstate of $k$ D1-branes 
suspended between the D3-branes in IIB string theory and the $k$-monopole solution 
in gauge theory was established in \cite{Diaconescu} by identifying the 
moduli space of the brane boundstates with the classical moduli space of the Nahm 
multi-monopole \cite{Nahm} in SYM. In addition, 
similarly to the instanton case before, the monopole gauge field itself can be
read off from the brane configuration using the brane-probe analysis.

This `brane within branes' realization of the monopole in string theory
is also confirmed/illustrated via a simple geometrical picture \cite{Hashimoto}.
Since in the vacuum the scalar field $\phi_1$ represents the separation
between the D3-branes, the scalar monopole field \eqref{scmonoun} 
should correspond to the deformation of the D3-branes pulled
by the monopole D1-string. Following \cite{Hashimoto} we show the D-monopole
profile by plotting \eqref{scmonoun} in Figure 6.
\begin{figure} [ht]
\label{fig6}
\begin{center}
{\scalebox{0.6}{
\includegraphics{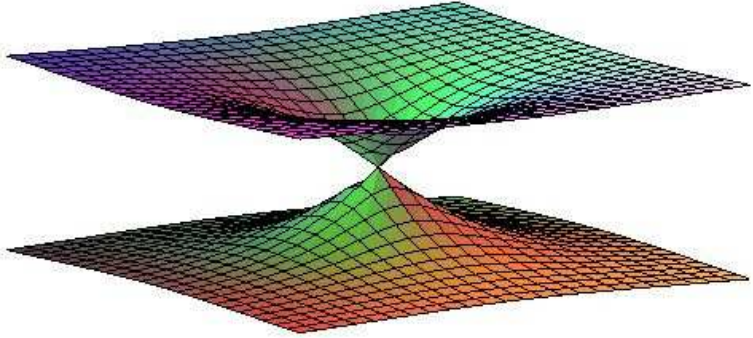}}
}
\end{center}
\caption{\small \it The monopole as the D1-string suspended between two D3-branes.
Its tension makes the D1-string shrink to a point and
at the same time pulls
the D3-branes together as two cusps which
are attached to the D-string.}
\end{figure}
The presence of the cusp where the two D3-branes meet
in this picture is interpreted as the D1-string suspended between the two D3-branes.
The D3-branes are pulled together by the tension of the D1-string stretched
between them, which shrinks to a point.

We can now compare this picture to the D-instanton within the separated
D3-branes. Since the gauge theory is now in the Coulomb phase, the vev $v$
is non-zero and we should use the constrained instanton solution \eqref{inst}.
Plotting the scalar field in \eqref{inst} in Figure 7 we find a
qualitatively different picture from Figure 6. There is no cusp at the 
point where the D3-branes meet. This is consistent with the fact that the 
D$(-1)$-brane is a point-like defect and not a string 
as in Figure 6.
\begin{figure} [ht]
\label{fig7}
\begin{center}
{\scalebox{0.8}{
\includegraphics{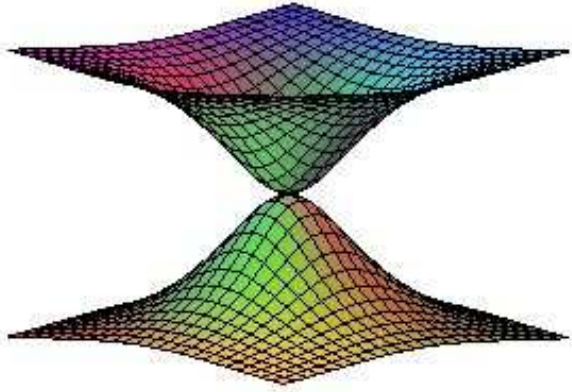}}
}
\end{center}
\caption{\small \it D-instanton as the D$(-1)$-brane placed between two D3-branes.
The D3-branes are smoothly deformed  and meet at the location
of the D$(-1)$.}
\end{figure}
For a general $p$ we see in Figure 8 that
the instanton is a D$p$-brane parallel to the D$(p+4)$-branes which is situated
exactly half way between them. The D$(p+4)$-branes are attracted by the instanton,
but the instanton worldvolume lies entirely along the larger branes. 
Hence, the instanton is a brane within branes and the monopole is a brane
stretched between branes.
(For simplicity we will continue referring to both
realizations as branes within branes.)
\begin{figure} [ht]
\label{fig8}
\begin{center}
{\scalebox{0.6}{
\includegraphics{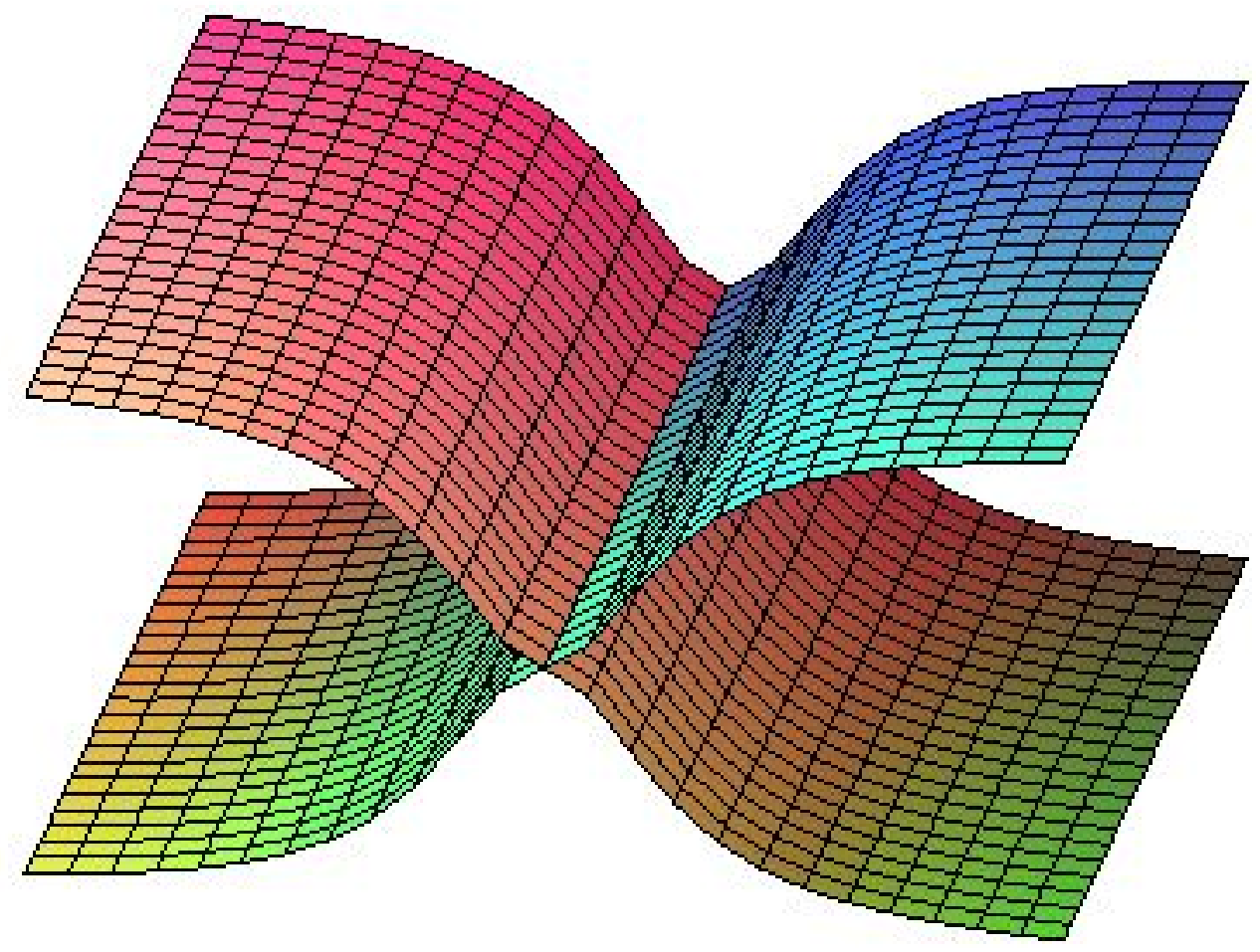}}
}
\end{center}
\caption{\small \it The instanton as the D$p$-brane within D$(p+4)$-branes.
}
\end{figure}

The worldvolumes of D3-branes are smoothly deformed by the instanton
between them with the curvature of the deformation determined by the
SYM instanton size $\rho$. 
It is also known that when $\rho \rightarrow 0$ the D$(-1)$-brane in
string theory can escape from the D3-branes \cite{Witten-i2}. 
This corresponds to a phase transition from the Higgs to the Coulomb phase
of the combined D$p$/D$(p+4)$ gauge theory.

The reader might ask whether the constrained nature of the SYM instanton
solution can affect the details of the instanton-brane picture on Figures 7
and 8.
The answer is simple: since the short-distance scalar field in \eqref{inst} 
is purely Abelian, and since the long-distance field is always Abelian, 
the scalar field in \eqref{inst} 
can be used at all distances,
due to Affleck's patching conditions 
\cite{Affleck} between the short- and the long-distance regimes.
It should also be added that in the string theory realization of the instanton,
in general there is no arbitrariness associated with a choice of Affleck
constraints, only the short-distance instanton is relevant for the
instanton partition function \cite{DHKM-rept}.

We note that the topology of all possible static stable
solutions on stacks of parallel D-branes
was studied exhaustively by Semenoff and Zarembo in \cite{Semenoff}. 
In the present paper we will need only the well-known examples
of such solutions -- the 
D-instantons and monopoles and their T-dual generalizations. 
As mentioned earlier, our
main goal is in studying time-dependent processes
involving solutions with opposite charges -- brane-antibranes --
and the appearance of lower-dimensional branes from brane-antibrane
pairs.

Finally, we want to comment on the relation between classical branes 
and their quantum excitations.
Classically, the monopole solution in gauge theory
represents a non-trivial vacuum.
Particle excitations appear from the lowest lying normal modes
when we first-quantize around this vacuum.
The eight fermion zero modes \eqref{lss} are essential in this 
set-up\footnote{Bosonic zero modes of the monopole 
also play a role. They are the 3-translations
in space and the $U(1)$ global rotations in the unbroken gauge group. 
When quantized, the latter give the tower of electrically charged dyons
which are important for the full $SL(2,Z)$ duality of the theory.
}
since their creation operators acting on the monopole vacuum fill in
precisely the right number of states to give a vector $\cN=4$ supermultiplet,
for a review see e.g. \cite{Harvey-lec}. In particular, there are precisely
as many particle states in the monopole supermultiplet as in the supermultiplet
of the W-boson $W^+$. This is one of the key elements \cite{Osborn} in
support of the electric-magnetic self-duality conjecture \cite{MO} of the
$\cN=4$ SYM.
When the $\cN=4$ SYM is realized as the worldvolume theory on D3-branes,
this electric-magnetic duality becomes a part of the S-duality of type IIB string
theory. In particular, monopoles (D1-strings suspended between D3-branes)
can be interchanged with electrically charged W-bosons (fundamental F1-strings
stretched between D3-branes). The D1-strings and the F1-strings 
are dual to each other as classical
extended objects; the duality between the monopole supermultiplet of states
and the supermultiplet of $W^+$ then arises from quantizing
the D1 and the F1 strings (with the usual difficulty that when the electric
modes are weakly coupled, the magnetic ones are strongly coupled and vice-versa).
The analysis in  this paper involves classical branes; their
particle excitations would arise from the first quantization 
and will not be relevant for the branes from branes programme we want to pursue.

\subsection{Imaginary time and real time processes through the sphaleron barrier}
 
Let us first consider the evolution in imaginary time of parallel 
(or coincident) D3-branes
passing through the D-instanton. The D-instanton is a D$(-1)$-brane
which is located at a point in space and time within the worldvolume
swept by the evolution of the 3-space-dimensional D3-branes,
as shown in Figure 1.
Before the encounter with the instanton, the worldvolume fields on the D3-brane
are in a vacuum state. Passing through the instanton, the fields on the D3-worldvolume
change, and at a late (Euclidean) time settle to another vacuum state, as illustrated
in Figure 9. The two classical vacua are topologically distinct, their
$S^3\rightarrow S^3$ winding numbers differ by one.  
This picture is a brane realization of the gauge theory instanton 
which describes the tunneling process between topologically distinct 
vacua.
\begin{figure} [ht]
\label{fig9}
\psfrag {Et} {\LARGE${\rm Imaginary\,time}$}
\psfrag {Mt} {\LARGE${\rm Real\,time}$}
\psfrag {Di} {\LARGE${\rm Instanton}$}
\psfrag {x13} {\LARGE${x_1,x_2,x_3}$}
\psfrag {t} {\LARGE${t}$}
\psfrag {x4} {\LARGE${x_4}$}
\begin{center}
{\scalebox{0.5}{
\includegraphics{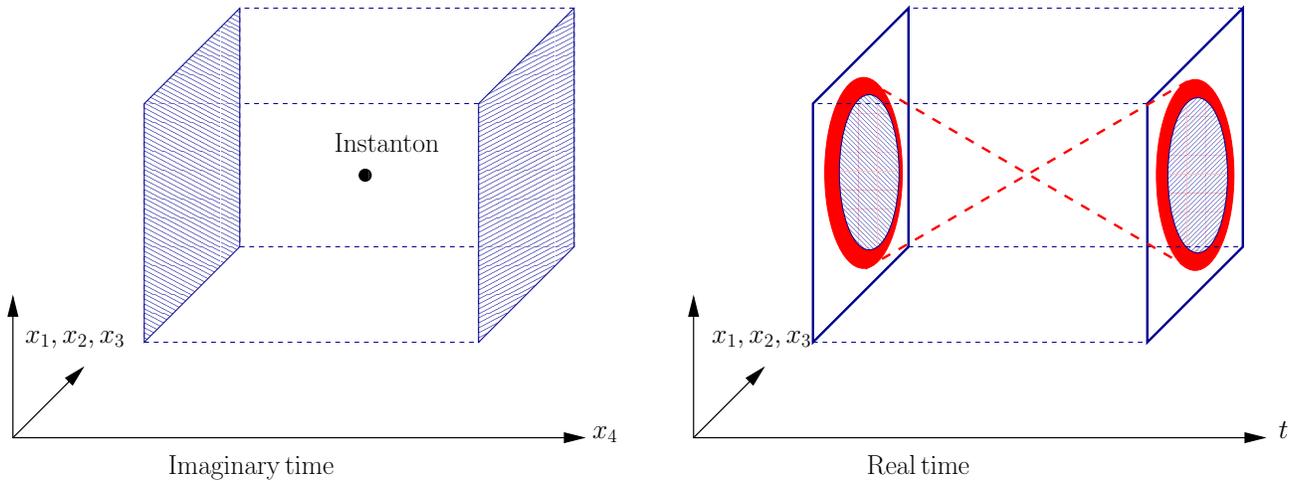}}
}
\end{center}
\caption{\small \it D$(-1)$ instanton brane inside the Euclidean worldvolume
of D3-branes versus a non-trivial Minkowski space solution. For clarity,
D3-branes are shown as coincident; the cube in each picture represents their
common worldvolume.
Vertical planes 
represent the D3-brane in the past and in the future time instances. Shaded regions
in each picture indicate topologically distinct vacuua
on the D3 brane. Ring-shaped regions in red, surrounding the 
vacua on the right picture depict
the radiation shell in the past and in the future.
}
\end{figure}
When vev is non-zero, there is a barrier between the vacua 
(as in Figure 1)
which corresponds to
the unstable classical solution in gauge theory, the sphaleron \cite{Manton}.

If one attempts to analytically continue the instanton \eqref{inst} 
to Minkowski space one encounters two immediate problems:
(1) the instanton becomes complex and
(2) it is not point-like anymore, but lives on the lightcone.

A better way to describe real time instanton-like processes is to look
for genuine Minkowski-space classical solutions which change the vacuum.
A class of such solutions with finite energy was
found in \cite{FKS}. These solutions describe spherical shells of radiation
first imploding, collapsing and then expanding in time. 
The region inside and outside the shell is a vacuum. As the solution evolves,
the vacuum {\it inside} the shell can change
to a topologically different vacuum 
(i.e. vacuum changes before and after the collapse
of the shell). 
This process inside the D3-branes is illustrated in the second picture
on Figure 9. The time-dependent solution lives on the light-cone (in the absence of 
vevs), in this way it resembles the Wick-rotated instanton, but there are no complexities
in genuine real-time solutions.  

When the $\cN=4$ theory is in the Coulomb phase one can fine-tune the incoming
classical radiation to pass precisely through (or close to) the top of the
spaleron barrier separating different vacua. In the infinite past one would start
with an imploding classical solution which at time zero will collapse to the
sphaleron. The spahleron will eventually decay and produce an exploding shell in
the infinite future. This process is illustrated in Figure 10 which is interpreted
as the production of an unstable D0 sphaleron brane in the type IIB theory.
\begin{figure} [ht]
\label{fig10}
\psfrag {x13} {\LARGE${x_1,x_2,x_3}$}
\psfrag {t} {\LARGE${t}$}
\psfrag {Sp} {\LARGE${\rm Sphaleron}$}
\begin{center}
{\scalebox{0.5}{
\includegraphics{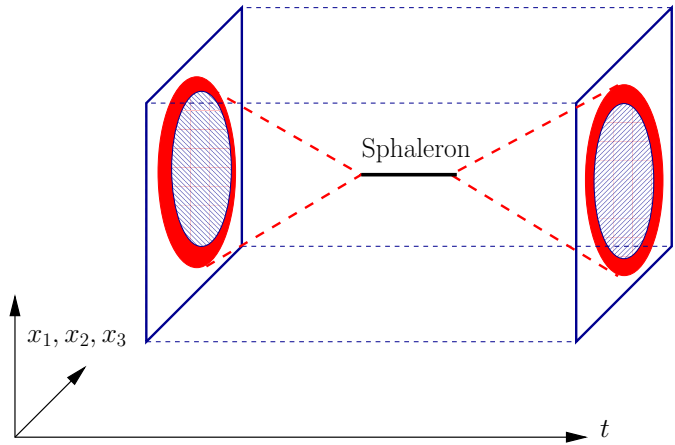}}
}
\end{center}
\caption{\small \it When the incoming radiation is fine-tuned to pass
through the top of the sphaleron barrier separating two topologically distinct 
vacua, the sphaleron solution is produced as the unstable D0-brane inside
the D3-worldvolume.
}
\end{figure}
By T-dualizing we can add common spatial dimensions to the D3 and the D0
world-volumes, producing unstable D$p$-branes with $p$ odd/even in type IIA/B theory.
Unstable D-branes in string theory have been identified with 
sphalerons in SYM already by Harvey, Horava and Kraus in \cite{Harvey-sp}.
These authors emphasized that this identification leads to a highly nontrivial
vacuum structure of string theory.

\subsection{Interpretation of the non-contractible monopole loop in imaginary time}

Returning now to the monopole-antimonopole loop of section {\bf 2.4},  
we recall 
that the loop parameter
$\tau$ should not be thought of as the real time $x_0$. 
The real time dimension will be {\it added}
to the loop in Section 3.

The non-contractible monopole loop is interpreted as a 
classical `process' 
of instanton creation in imaginary time $\tau=x_4$. More precisely
this loop is a blow-up of the instanton which shows the instanton constituents.
Since the instanton is a (Euclidean) time-dependent configuration, it should be thought of
as a `process' in $x_4$ rather than a `particle'. The monopole loop is the instanton
and it shows that the instanton `process' is made out of constituent monopole particles
which are first created from the vacuum
and then annihilate each other in imaginary time. 
It was suspected for a long time \cite{instquarks}
that in gauge theories instantons should be thought of
as composite states of more basic configurations referred to as `instanton quarks'.
We conclude that the monopoles are
the instanton quarks.  

In a context, when one of the dimensions is finite,
this conclusion was tested via explicit calculations of the gluino condensate
in our earlier work \cite{DHKM-mono,DHK-mono}.
Note that there is an alternative way to construct a non-contractible
monopole loop when there is a compact dimension. 
Instead of introducing the Taubes winding by $U(1)$-gauge rotating
the monopole, one can wind the monopole worldline along a compact direction
\cite{Nahmtwo,Yi-cyl,KvBaal,DHKM-mono}. Such a non-contractible monopole loop also gives
rise to an instanton solution in agreement with Taubes arguments. More concretely,
in \cite{Yi-cyl} it was verified that the instanton on partially compactified D-branes
is a composite configuration made of monopoles with the net magnetic charge zero
and with one unit of winding along the compactified worldvolume direction.
In \cite{KvBaal} the periodic instanton in high-temperature QCD was identified 
as a composite monopole-antimonopole configuration.
In Ref. \cite{DHKM-mono} the instanton solution in 
$\cN=1$ pure SYM on $R^3 \times S^1$ was decomposed into its constituents:
magnetic monopoles (with unit net winding around $S^1$ and vanishing net magnetic
charge). On $R^3 \times S^1$ these monopoles have finite Euclidean action,
since the dimension compactified on $S^1$ is finite.
These monopoles are the elementary 
semiclassical configurations contributing to the path integral. 
The gluino condensate can be calculated exactly on these monopole configurations, and
the results \cite{DHKM-mono,DHK-mono} are 
in complete agreement with the known values\footnote{Meaning the correct values
obtained in the weakly coupled theory -- the WCI results -- see 
\cite{DHKM-mono} for more detail.}
of the gluino condensate for all classical
gauge groups. 

Our SYM construction of the non-contractible monopole-antimonopole loop is embedded
into type IIB string theory
in an obvious way following the discussion in the previous subsection.
It is realized as the D1-brane--D1-antibrane non-contractible loop suspended
between two D3-branes.
The worldvolumes of the D3-branes are Euclidean i.e. the D3-branes are S-branes
\cite{S-branes}
spanning the $x_1,\ldots,x_4$ purely spatial dimensions. 
The loop parameter is identified with the spatial dimension $x_4$ , hence, 
 D1-branes are
also S-branes. This D1-brane--D1-antibrane non-contractible loop within
two D3-branes is sketched in Figure 11.
By minimizing the action of the brane-configurations in Figure 12
we obtain the D$(-1)$-brane within the D3-branes -- the string instanton.
\begin{figure} [ht]
\label{fig11}
\psfrag {Db} {\LARGE${\overline{\rm D}1-{\rm brane}}$}
\psfrag {D} {\LARGE${{\rm D}1-{\rm brane}}$}
\psfrag {x4} {\LARGE${x_4}$}
\psfrag {x3} {\LARGE${x_3}$}
\psfrag {x12} {\LARGE${x_1,x_2}$}
\begin{center}
{\scalebox{0.5}{
\includegraphics{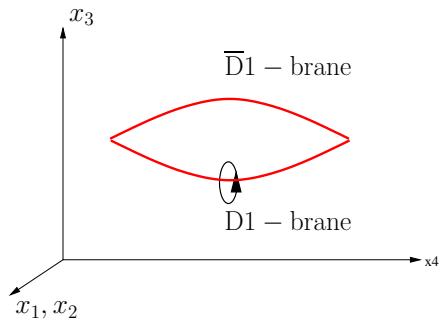}}
}
\end{center}
\caption{\small \it The D1-brane--D1-antibrane
 non-contractible loop as seen from the worldvolume of one of the D3-branes.
 Red lines trace the ends of the D1 and anti-D1 strings as they evolve
 in $x_4$. The circular arrow denotes the $U(1)$ twist of the D1-brane
 in the D3-brane worldvolume. 
}
\end{figure}

\section{D$(p-2)$-branes from D$p$-brane--D$p$-antibrane annihilation}

\subsection{D0-brane from D2-brane--D2-antibrane annihilation}

Now we are ready to add a real time dimension to the 
purely Euclidean  considerations of the previous section.
Let us consider the deformation of Figure 11 depicted on Figure 12.
This Figure can be seen as a sequence of snap-shots in time starting at
$t=0$ with Figure 11, and evolving backwards in time to the far
separated D2-brane and D2-antibrane at $t=-T_0$. The Euclidean time
evolution of the monopole-antimonopole configuration of the previous section
is now replaced by the extend of D2-branes in the spatial $x_4$ direction.
\begin{figure} [ht]
\label{fig12}
\psfrag {Db} {\LARGE${\overline{\rm D}2}$}
\psfrag {D} {\LARGE${{\rm D}2}$}
\psfrag {x4} {\LARGE${x_4}$}
\psfrag {x0} {\LARGE${t}$}
\psfrag {-T0} {\LARGE${t=-T_0}$}
\psfrag {0} {\LARGE${t=0}$}
\begin{center}
{\scalebox{0.5}{
\includegraphics{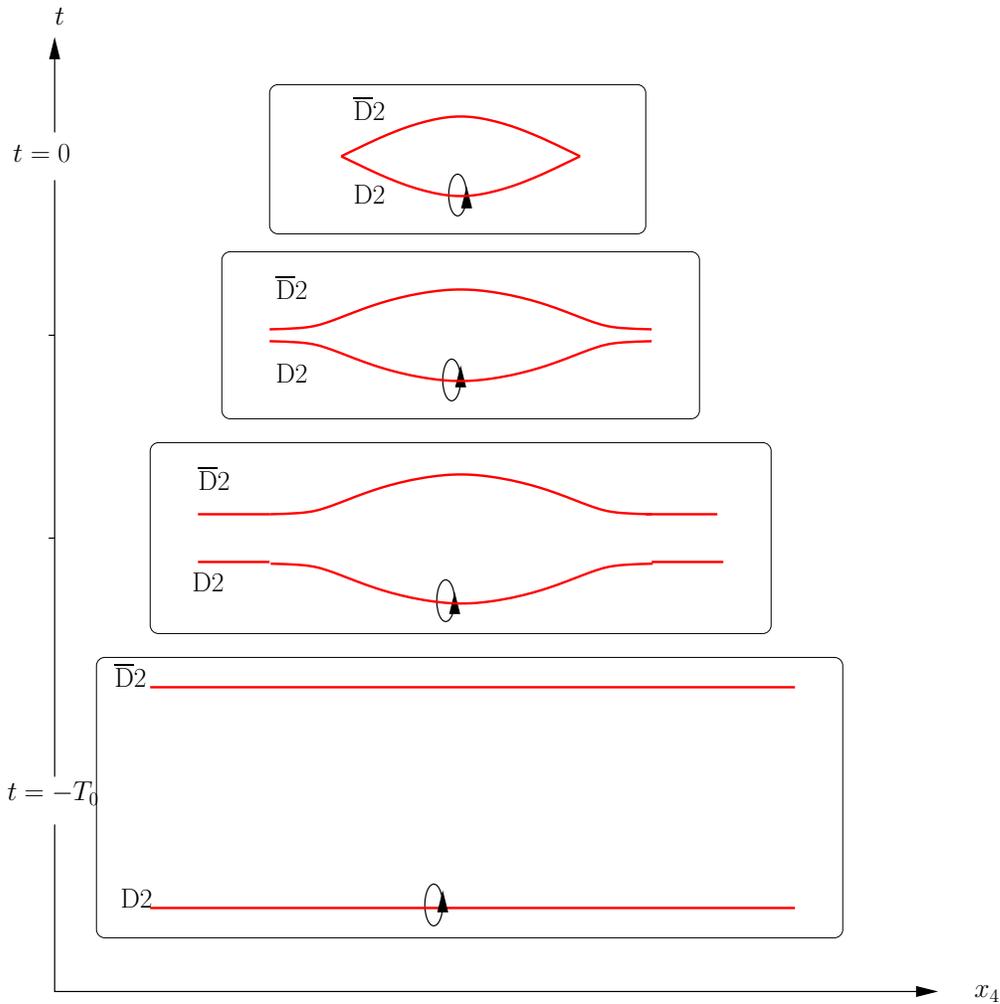}}
}
\end{center}
\caption{\small \it Annihilation of the D2-anti-D2 brane
configuration suspended between D4-branes. Due to the $U(1)$ twist,
denoted by the circular arrow, this annihilation process leads to a
formation of the D0-brane at $t>0$.}
\end{figure}

The time-dependent process we want to consider is the annihilation
of the D2- and the anti-D2-branes suspended between two fixed D4-branes.
At an early time, $t=-T_0,$ the 2-branes describe two parallel two-dimensional
surfaces in $(x_4,x_9)$ at a large separation from each other along, say, 
$x_3$. The two D4-branes span three spatial dimensions $(x_1,x_2,x_3)$
and are located at a fixed distance $2 \pi \alpha' v$ away from each other 
along $x_9$. We further require that this initial configuration is prepared
in such a way that one of the D2-branes is gauge-rotated in the D4-worldvolume
by a U(1) gauge transformation $U(x_4)$ with the winding number one.\footnote{
By winding number $n$
 we mean $U(x_4)=\exp[i\Lambda(x_4) \tau^3/2]$, with 
 $\Lambda(\infty) - \Lambda(0) =2\pi n$.}
This gauge twisting cannot be removed with a global $U(1)$ gauge transformation
in the D4 worldvolume as it corresponds to a {\it relative} gauge orientation
factor of the D2-anti-D2 pair.

Evolving this initial configuration forward in time, the D2-brane annihilates
the D2-antibrane at $t=0$, and leaves behind (amid perturbative radiation)
a D0-brane whose worldline is along the $t\geq 0$ ray in the COM frame
of the collision as in Figure 8.  The D0-worldline is parallel
to the D4-worldvolumes, it is the instanton of subsection {\bf 2.5}.
The D0-brane appears as a topological
soliton in the worldvolume theory of the D4-branes from the process
of the D2-anti-D2 annihilation. The RR-charge of this D0-brane is the winding number
of the gauge twist of the D2-anti-D2 pair.

In the approach of Sen \cite{Sen} the D0-brane is the topological soliton
or kink of the complex tachyon field which lives on the
worldvolume of the D2 brane-antibrane pair. The appearance and stability
of the kink depend crucially on the hypothesis of the tachyon condensation and on the
form of the conjectured tachyon potential. In our approach we choose to work on the
worldvolume of the external D4-branes, and the tachyon, which lives on the
worldvolume of the D2-anti-D2 pair, does not make an appearance. As already stated
earlier, the D0-brane charge in our approach originates from the D2 gauge twisting
on the D4-worldvolume.

In general, it is clear that if one requires the infinite ends of the 
D2 and the anti-D2 branes to completely annihilate
each other at time $t>0$,
the $U(1)$ gauge twisting must have an integer (or vanishing) winding number.
Another way to think about it is to imagine the $x_4$ dimension being compactified.
If one does not wish to impose such a restriction,
it is natural to expect that as $t \rightarrow 0$ and the D2-branes approach
each other, a generic gauge twisting will cluster into a product of gauge transformations,
each with a support in a local region and an integer winding number in this region.
D0 and anti-D0 branes will be produced locally in these
local regions with the charges
prescribed by the local winding numbers. The ends of the D2-branes at infinity
(in $x_4$) 
would not match and will live some radiation debris.

\subsection{Generalization to D$p$-brane--D$p$-antibrane pairs suspended between 
D$(p+2)$-branes}

The treatment of higher dimensional cases is a straightforward generalization
of the picture developed in the previous subsection. It proceeds by adding $p-2$ common
spatial dimensions to all of the worldvolumes of  the branes and antibranes
involved. 

The resulting process in the type II string theory
is a mutual annihilation of a D$p$-brane
with a D$p$-antibrane suspended between two D$(p+2)$-branes and with a non-trivial
$U(1)$ net winding along the $x_4$ direction. After the annihilation
one is left
with D$(p-2)$-branes within the spectator D$(p+2)$-branes.
The D$(p-2)$-charges are determined by the $U(1)$ winding
in the D$(p+2)$-worldvolume.

The highest dimension we can address in this way is dictated by the
dimensionality of the  D$(p+2)$-branes.
In type IIA theory we can have a D6-anti-D6 pair suspended between two D8-branes and
leading to stable D4-branes within the same spectator branes. We denote this process as,
\be
{\rm IIA:}  \qquad {\rm D}8 \ - \ \big[{\rm D}6-\overline{\rm D}6\big]
\ - \ {\rm D}8 \
\longrightarrow \  {\rm D}8 \ - \ \big[{\rm D}4\big] \ - \ {\rm D}8 \ . \label{des8}
\ee
By successive applications of T-duality in type II string theory 
(or by dimensional reduction in the
9-dimensional gauge theory) we can access all the lower cases up to, 
\be
{\rm IIB:}  \qquad {\rm D}3 \ - \ \big[{\rm D}1-\overline{\rm D}1\big]
\ - \ {\rm D}3 \
\longrightarrow \  {\rm D}3 \ - \ \big[{\rm D}(-1)\big] \ - \ {\rm D}3 \ . \label{des3}
\ee
Before we list classify the brane descent relations obtained 
in this way we want to pause to examine the process which is
S-dual to the last equation,
\be
{\rm IIB:}  \qquad {\rm D}3 \ - \ \big[{\rm F}1-\overline{\rm F}1\big]
\ - \ {\rm D}3 \
\longrightarrow \  {\rm D}3 \ - \ \big[\tilde{\rm D}(-1)\big] \ - \ {\rm D}3
 \ . \label{des3S}
\ee

\subsection{S-duality in IIB and the S-dual of the D-instanton}

The type IIB superstring theory is believed to be invariant under 
$SL(2,Z)$ duality transformations \cite{125,127,110}. We will need to consider here
only the S-duality generator of this group which interchanges
the fundamental string, F1, with the D1-string. The fate of other branes
in type IIB is as follows:
the D3-brane is mapped to itself and the D5-brane is interchanged with the solitonic 
NS-5 brane,
\bea 
&&{\rm D}1 \  \longrightarrow \ \tilde{\rm D}1 \ = \ {\rm F}1 \ , \qquad
{\rm F}1 \  \longrightarrow \ \tilde{\rm F}1 \ = \ {\rm D}1 \ , \\
&&{\rm D}3 \  \longrightarrow \ \tilde{\rm D}3 \ = \ {\rm F}3 \ , \\
&&{\rm D}5 \  \longrightarrow \ \tilde{\rm D}5 \ = \ {\rm NS}5 \ , \qquad
{\rm NS}5 \  \longrightarrow \ \tilde{\rm NS}5 \ = \ {\rm D}5 \ .
\eea
One way to derive these relations is by calculating tensions of these branes,
\be
\tau_{\rm Dp} \,=\, {1 \over (2\pi)^p \,{\alpha'}^{p+1\over 2} g_{\rm st}} \quad, \quad
\tau_{\rm F1} \,=\, {1 \over 2\pi{\alpha'}} \quad, \quad
\tau_{\rm NS5} \,=\, {1 \over (2\pi)^5\, {\alpha'}^{3} g_{\rm st}^2} \ ,
\label{tens}
\ee    
and using the S-duality dictionary
\be
g_{\rm st}\ \rightarrow \ \tilde{g}_{\rm st} \,=\, {1 \over g_{\rm st}} \quad , \quad
\alpha' \ \rightarrow\ \tilde{\alpha'} \,=\,\alpha'g_{\rm st} \ ,
\label{dict}
\ee 
to equate them, see e.g. \cite{PolchinskiII,CVJ}.

 An interesting question to ask is what is the S-dual of the D$(-1)$-brane (and similarly
 of its magnetic dual D7 brane). We will argue now that the S-dual of
 the D-instanton is a new $(-1)$-brane in the type IIB theory, which we denote as 
 $\tilde{\rm D}(-1)$ (and similarly there is a second 7-brane $\tilde{\rm D}$7).
 First, it is obvious from the first equation in \eqref{tens} and the dictionary
 \eqref{dict} that the tensions of the $\tilde{\rm D}(-1)$-brane,
 and of the $\tilde{\rm D}$7 brane are different from the D$(-1)$ and the D7 tensions,
 \bea \label{dinten}
 \tau_{\tilde{\rm D}(-1)} \,=\, {2\pi \over \tilde{g}_{\rm st}} &=& 2\pi  g_{\rm st} \quad ,
 \quad
 \tau_{{\rm D}(-1)}\,=\, {2\pi \over  g_{\rm st}} \ , \\
 \tau_{\tilde{\rm D}7} \,=\, {1 \over (2\pi)^7 \, \tilde{\alpha'}^4 \tilde{g}_{\rm st}}
 &=& {1 \over (2\pi)^7 \, {\alpha'}^4 {g}_{\rm st}^3}
  \quad ,
 \quad 
 \tau_{{\rm D}7}\,=\, {1 \over (2\pi)^7 \, {\alpha'}^4 {g}_{\rm st}} \ .
 \eea
It also appears that the S-dual of the instanton, the $\tilde{\rm D}(-1)$-brane,
has a `perturbative' tension $\propto {g}_{\rm st}$ in terms of the parameters of
the original theory. Are there really two types of $(-1)$-branes in type IIB?

If we look at the D-instanton as a classical supergravity solution \cite{Dinst},
we discover that it is mapped to itself under the S-duality transformation 
\be
\tau(x) \, \longrightarrow \, -{1\over \tau(x)} \quad, \quad
{\rm where} \quad
\tau(x) \, = \, C^{(0)}(x) +\, i\,e^{-i\phi(x)} \ .
\label{strans}
\ee
In other words, it appears that there is only one type of D-instanton solution
in type IIB supergravity. Technically, the instanton components of the
dilaton, $\phi(x)$, and of the 
RR-scalar field, $C^{(0)}(x)$, change non-trivially under \eqref{strans},
but when they are combined into $\tau(x)$, simplifications occur.
The instanton solution in \cite{Dinst} is constructed in such a way that
the complexified scalar field $\tau(x)$ is actually a constant,
$\tau_{\rm D-inst}(x)=\langle\tau \rangle,$ and the spacetime-dependent contributions
in $\phi(x)$ and $C^{(0)}(x)$ cancel each other.
 Hence, there is only one instanton
solution in supergravity, but under the S-duality transformation, the asymptotic
value of $\tau(x)$ changes, 
\be
\langle\tau \rangle \, \longrightarrow\,
\langle\tilde{\tau} \rangle \,  =
-{1\over \langle\tau\rangle} \ .
\label{taurel}
\ee
S-duality is not a symmetry of the theory. Under the transformation \eqref{strans} 
one supergravity formulation goes to a different one. Both formulations have the same
Lagrangian, but different values of $\langle\tau \rangle$ which are related via
\eqref{taurel}. These two theories describe the same physics in terms of different
degrees of freedom, $\tau(x)$ in the first version, and $\tilde{\tau}(x)$  in the
S-dual version. There is an instanton solution in each of these theories,
which has the same algebraic form when expressed in terms of the fundamental
fields of each theory. But 
the D-instanton and its S-dual are different objects with
different tensions, in agreement with \eqref{dinten}. 

Exactly the same conclusion is reached in the $\cN=4$ SYM case (which of course is 
another simple limit of the type IIB string theory). There is an instanton solution
\eqref{inst} in the original formulation of the theory in terms of $A_\mu$ and
$\phi$. The S-dual formulation of this theory is in terms of $\tilde{A}_\mu$ and
$\tilde\phi$, which describe monopole degrees of freedom. This theory too has the
instanton solution \eqref{inst} but in terms of $\tilde{A}_\mu$ and
$\tilde\phi$. The actions of these two solutions are related again via \eqref{dinten}. 

We get an insight into the nature of the second instanton by S-dualizing
the original instanton when it is viewed as a composite configuration of monopoles
or, equivalently D1-anti-D1 branes. The S-dual of the instanton
is then the F1-anti-F1 configuration suspended between two self-dual D3-branes
(with the $U(1)$-twist),
\be
{\rm IIB:}  \qquad {\rm D}3 \ - \ \big[{\rm F}1-\overline{\rm F}1\big]
\ - \ {\rm D}3 \
\longrightarrow \  {\rm D}3 \ - \ \big[\tilde{\rm D}(-1)\big] \ - \ {\rm D}3
 \ . \label{des3Sagain}
\ee

Since our approach is classical in nature, we think of fundamental strings
F1 here as classical solutions to the Born-Infled action in the $\cN=4$ gauge theory,
the BIons \cite{47,46,48}. The classical BIons are S-dual to classical monopoles as
F1-strings are S-dual to D1-strings. 
We conclude that there is an S-dual instanton,
$\tilde{\rm D}(-1)$,  made of dual instanton quarks, which are  $W^+$ and
$W^-$ bosons.

\subsection{Stable brane descent relations in type II}

By successive applications of T-duality of type II theory
to \eqref{des8} 
and by using S-duality in type IIB theory as in the previous section,
we can have a variety of processes describing the production of stable
branes. These processes are summarized in Table 1.
\begin{table}%\setlength{\extrarowheight}{5pt}
\begin{center}\begin{tabular}{ccccc} \hline\hline
\phantom{$\Biggr($} spectator branes & brane-antibrane & lower brane\\
 \hline
 & & \\
 $2 \times {\rm D}8$ & ${\rm D}6-\overline{\rm D}6$ & ${\rm D}4$\\
 $2 \times {\rm D}7$ & ${\rm D}5-\overline{\rm D}5$ & ${\rm D}3$\\
 $2 \times \tilde{\rm D}7$ & ${\rm NS}5-\overline{\rm NS}5$ & ${\rm D}3$\\
 $2 \times {\rm D}7$ & ${\rm NS}5-\overline{\rm NS}5$ & ${\rm D}3$\\
 $2 \times \tilde{\rm D}7$ & ${\rm D}5-\overline{\rm D}5$ & ${\rm D}3$\\
 $2 \times {\rm D}6$ & ${\rm D}4-\overline{\rm D}4$ & ${\rm D}2$\\
 $2 \times {\rm D}5$ & ${\rm D}3-\overline{\rm D}3$ & ${\rm D}1$\\
 $2 \times {\rm NS}5$ & ${\rm D}3-\overline{\rm D}3$ & ${\rm F}1$\\
 $2 \times {\rm D}4$ & ${\rm D}2-\overline{\rm D}2$ & ${\rm D}0$\\
 $2 \times {\rm D}3$ & ${\rm D}1-\overline{\rm D}1$ & ${\rm D}(-1)$\\
 $2 \times {\rm D}3$ & ${\rm F}1-\overline{\rm F}1$ & $\tilde{\rm D}(-1)$\\
 & & \\
\hline\hline
\end{tabular}\end{center}
\caption{\small Brane-antibrane annihilations to lower-deimensional branes.}
\end{table}

The brane descent relations in Table 1 are in agreement with those obtained 
from Sen's tachyon condensation approach \cite{Sen,Witten-k,Sen-d}, 
except that we cannot describe annihilations
of D9, D8 and D7 branes.
The highest dimension we can address is restricted by the 
dimensionality of spectator branes. 

It might be tempting to dispose
of the spectator branes altogether, but then we 
would not be able to use the SYM language
of their worldvolume theory to describe annihilations of lower-dimensional branes.
The spectator branes are certainly not necessary for brane-antibrane annihilations
to occur and for smaller branes to be produced in string theory. One would just
need to use a different approach to describe this -- 
the original tachyon condensation conjecture of Sen
\cite{Sen}. Our approach provides a complimentary
picture to \cite{Sen}. Using our method we are also able to gain insight 
into non-Dirichlet stable branes and the S-dual of the D-instanton.

\section{Comments and open questions}

We explained how stable $q$- and unstable $(q+1)$-branes in string theory appear from
$(q+2)$-brane-antibrane pairs in the background of two 
stable $(q+4)$-branes
(here $q$ is even/odd for type IIA/B theory). 
These lower-dimensional branes appear as classical configurations
in the SYM worldvolume theory on the D$(q+4)$-branes.

More specifically, we argued that stable $q$-branes are produced in
annihilation processes of $(q+2)$-branes with $(q+2)$-antibranes, and that
their RR charge is carried by the $U(1)$ winding of the $(q+2)$-brane in the 
$(q+4)$-brane worldvolume. The $q$-branes are stable BPS configurations
and their RR charges are under control.\footnote{It is known 
\cite{Douglas-i2} that the RR-charge of the D$q$-brane within D$(q+4)$-branes
coincides with the instanton charge \eqref{Qinst}, and that the latter 
does appear from the non-contractible monopole-antimonopole loop \cite{Taubes-lect} as
discussed in subsection {\bf 2.4}.}
The string theoretical interpretation of the $U(1)$ winding is as follows:
the $U(1)$ gauge transformation of the $(q+2)$-branes corresponds to a rotation of the
strings stretching between the $(q+2)$-brane and the two $(q+4)$-branes, such that 
the strings ending on the upper and the lower $(q+4)$-brane are rotated in opposite
directions.
What needs to be understood better is the
relation of this $U(1)$-winding  to the winding of the complex tachyon field
in the `Mexican Hat' tachyon potential proposed by Sen \cite{Sen}. 

Tachyonic modes on the $(q+2)$-brane--$(q+2)$-antibrane
worldvolume do not appear in our analysis since we work on the worldvolume
of the spectator $(q+4)$-branes. The forces between $(q+2)$-branes and 
$(q+2)$-antibranes at large separations are the standard brane-antibrane
forces, as in \cite{BS}, and in the presence of the D$(q+4)$ spectator branes
these forces are fully reproduced in gauge theory on the D$(q+4)$-worldvolume.

At small separations of the $(q+2)$-brane--$(q+2)$-antibrane the same physics 
can be described in two very different languages. 

(1) In the language of tachyon condensation \cite{Sen},
a new tachyon channel opens up between the brane and the antibrane and 
the tachyon condensation describes the annihilation of the brane-antibrane
pair into the true vacuum.
The unstable configuration at the top of the
tachyon potential corresponds to the coincident brane-antibrane configuration,
and the ground state is the closed string vacuum. Lower-dimensional branes appear
as solitons in the tachyon field.

(2) The second language is the gauge theory on the spectator D$(q+4)$-branes
used in this paper. All other lower-dimensional branes are realized
as classical solutions in this theory.
At small separations, the fields representing the $(q+2)$-brane
and the $(q+2)$-antibrane
 start eating each other leading to a destruction of
the $(q+2)$-brane-antibrane configuration. If the $U(1)$-winding is trivial,
the destruction is complete and one ends up in the vacuum. For a non-vanishing
$U(1)$-winding, one is left with topological solitons describing
stable D$q$-branes within the spectator D$(q+4)$-branes.

In the tachyon condensation approach the brane and the antibrane 
are not modified even at zero separations. Instead, the tachyon channel opens
up and the configuration becomes unstable. In the branes within branes approach,
there are no tachyons, but the brane and the antibrane partially or completely
annihilate each other as classical objects. It would be very interesting to
understand better the precise relation between these two approaches.

Finally, we would like to comment on unstable branes. Unstable branes 
arise in our approach as unstable classical solutions in the 
worldvolume gauge theory.
It should be clear from sections {\bf 2.3} and {\bf 2.4} that these
sphaleron  D$(q+1)$-branes are directly related to both: the
monopole D$(q+2)$-branes, and the instanton D$q$-branes. The existence of 
unstable D$(q+1)$-branes follows from the minimax procedure applied
to the non-contractible D$(q+2)$-loop, and they represent
the saddle-point configurations  on top of the barrier under which the
Euclidean D$q$-branes tunnel. Whenever D-instanton S-branes
are present, there exists a finite-energy real-time process
which produces a sphaleron solution of one dimension higher as in Figure 10.
Stable D-branes can be produced in real-time collisions of higher branes
as explained in section {\bf 3} and in Figure 12. 
However, unstable sphaleron-branes
do not appear directly in these processes. Unstable branes 
in string theory have been
identified with the SYM sphaleron solutions already in 
\cite{Harvey-sp,Gross-sp}, but as these branes carry no RR charges
and are not BPS-protected, there is not much we can infer from the SYM side
about them, apart from their existence.

The derivation of descent relations between the stable branes summarized in section
{\bf 3.4} is one of the main results of this paper. These relations are in agreement
with general K-theory considerations \cite{Witten-k}. Two features of these
relations are particularly interesting.
First, is that the same  $(q+2)$-brane-$(q+2)$-antibrane pair can produce 
different $q$-branes depending on the type of spectator branes used as in
\bea
&{\rm D}5 \ - \ \big[{\rm D}3-\overline{\rm D}3\big]
\ - \ {\rm D}5 \
&\longrightarrow \  {\rm D}5 \ - \ \big[{\rm D}1\big] \ - \ {\rm D}5  \nonumber\\
&{\rm NS}5 \ - \ \big[{\rm D}3-\overline{\rm D}3\big]
\ - \ {\rm NS}5 \
&\longrightarrow \  {\rm NS}5 \ - \ \big[{\rm F}1\big] \ - \ {\rm NS}5   \nonumber
\eea
It is not clear what would distinguish between these two processes
in the absence of the spectator branes, and how
the second process would arise in Sen's approach.

The second interesting feature is the appearance of the S-dual of the instanton
(and also of its magnetic dual)
as in  
\bea
&{\rm D}3 \ - \ \big[{\rm D}1-\overline{\rm D}1\big]
\ - \ {\rm D}3 \
&\longrightarrow \  {\rm D}3 \ - \ \big[{\rm D}(-1)\big] \ - \ {\rm D}3   \nonumber\\
&{\rm D}3 \ - \ \big[{\rm F}1-\overline{\rm F}1\big]
\ - \ {\rm D}3 \
&\longrightarrow \  {\rm D}3 \ - \ \big[\tilde{\rm D}(-1)\big] \ - \ {\rm D}3
  \nonumber
\eea
and its interpretation in {\bf 3.3} as a non-contractible BIon-anti-Bion loop.
The S-dual instanton $\tilde{\rm D}(-1)$ is
a point-like object with a perturbative action.
Though a classical solution in the dual theory, it would be interesting to find
its fully quantum interpretation in the original theory.

\bigskip

\centerline{\bf Acknowledgements}

I would like to thank the Aspen Centre for Physics and CERN Theory Division 
for hospitality. I acknowledge useful discussions with Adi Armoni, 
Jose Fernandez Barbon and Gabriele Travaglini.
I am grateful to Catherine and Nicholas for help, support and understanding
throughout.

%\bigskip

%%%%%%%%%%%%%%%%%%%%%%%%%%%%%%%%%%%%%%%%%%%%%%%%%%%%

%%%%%%%%%%%%%%%%%%%%%%%%%%%%%%%%%%%%%%%%%%%%%%%%%%

%%%%%%%%%%%%%%%%%%%%%%%%%%%%%%%%%%%%%%%%%%%%%%%%%%%%%%%%%%%


\newpage

\begin{thebibliography}{99}  
%\baselineskip 0pt  

\bibitem{Sen1}
A.~Sen,
``Stable non-BPS bound states of BPS D-branes,''
JHEP {\bf 9808} (1998) 010
[arXiv:hep-th/9805019].
%%CITATION = HEP-TH 9805019;%%
  
\bibitem{Sen}
A.~Sen,
``Tachyon condensation on the brane antibrane system,''
JHEP {\bf 9808} (1998) 012
[arXiv:hep-th/9805170];
%%CITATION = HEP-TH 9805170;%%
\\
``Non-BPS states and branes in string theory,''
arXiv:hep-th/9904207.
%%CITATION = HEP-TH 9904207;%%

\bibitem{Sen-d}
A.~Sen,
``Descent relations among bosonic D-branes,''
Int.\ J.\ Mod.\ Phys.\ A {\bf 14} (1999) 4061
[arXiv:hep-th/9902105].
%%CITATION = HEP-TH 9902105;%%

\bibitem{Sen3}
A.~Sen,
``Universality of the tachyon potential,''
JHEP {\bf 9912} (1999) 027
[arXiv:hep-th/9911116].
%%CITATION = HEP-TH 9911116;%%

\bibitem{BS} 
T.~Banks and L.~Susskind,
``Brane - Antibrane Forces,''
arXiv:hep-th/9511194.
%%CITATION = HEP-TH 9511194;%%

\bibitem{Witten-k} 
E.~Witten,
``D-branes and K-theory,''
JHEP {\bf 9812} (1998) 019
[arXiv:hep-th/9810188].
%%CITATION = HEP-TH 9810188;%%

\bibitem{Yietal} 
P.~Yi,
``Membranes from five-branes and fundamental strings from Dp branes,''
Nucl.\ Phys.\ B {\bf 550} (1999) 214
[arXiv:hep-th/9901159];
%%CITATION = HEP-TH 9901159;%%
\\
O.~Bergman, K.~Hori and P.~Yi,
``Confinement on the brane,''
Nucl.\ Phys.\ B {\bf 580} (2000) 289
[arXiv:hep-th/0002223].
%%CITATION = HEP-TH 0002223;%%

\bibitem{HTaylor} 
K.~Hashimoto and W.~Taylor,
``Strings between branes,''
JHEP {\bf 0310} (2003) 040
[arXiv:hep-th/0307297].
%%CITATION = HEP-TH 0307297;%%

\bibitem{Taubes-lect} 
C.~H.~Taubes,
``Morse Theory And Monopoles: Topology In Long Range Forces,''
In Cargese 1983, Proceedings, {\it Progress In Gauge Field Theory}, 
eds. G. 't Hooft et al, Plenum Press, New York 1984, pp. 563-587. 

\bibitem{Taubes} 
C.~H.~Taubes,
``The Existence Of A Nonminimal Solution To The SU(2) Yang-Mills Higgs Equations On R**3,''
Commun.\ Math.\ Phys.\  {\bf 86} (1982) 257; Commun.\ Math.\ Phys.\  {\bf 86} (1982) 299. 
%%CITATION = CMPHA,86,257;%%


\bibitem{Thooft-m} 
G.~'t Hooft,
``Magnetic Monopoles In Unified Gauge Theories,''
Nucl.\ Phys.\ B {\bf 79} (1974) 276.
%%CITATION = NUPHA,B79,276;%%

\bibitem{Polyakov} 
A.~M.~Polyakov,
``Particle Spectrum In Quantum Field Theory,''
JETP Lett.\  {\bf 20} (1974) 194
[Pisma Zh.\ Eksp.\ Teor.\ Fiz.\  {\bf 20} (1974) 430].
%%CITATION = JTPLA,20,194;%%

\bibitem{BPS} 
E.~B.~Bogomolny,
``Stability Of Classical Solutions,''
Sov.\ J.\ Nucl.\ Phys.\  {\bf 24} (1976) 449
[Yad.\ Fiz.\  {\bf 24} (1976) 861];
%%CITATION = SJNCA,24,449;%%
\\
M.~K.~Prasad and C.~M.~Sommerfield,
``An Exact Classical Solution For The 'T Hooft Monopole And The Julia-Zee Dyon,''
Phys.\ Rev.\ Lett.\  {\bf 35} (1975) 760.
%%CITATION = PRLTA,35,760;%%

\bibitem{Nahm} 
W.~Nahm,
``A Simple Formalism For The Bps Monopole,''
Phys.\ Lett.\ B {\bf 90} (1980) 413.
%%CITATION = PHLTA,B90,413;%%


\bibitem{BPST} 
A.~A.~Belavin, A.~M.~Polyakov, A.~S.~Shvarts and Y.~S.~Tyupkin,
``Pseudoparticle Solutions Of The Yang-Mills Equations,''
Phys.\ Lett.\ B {\bf 59} (1975) 85.
%%CITATION = PHLTA,B59,85;%%

\bibitem{Thooft-i} 
G.~'t Hooft,
``Computation Of The Quantum Effects Due To A Four-Dimensional  Pseudoparticle,''
Phys.\ Rev.\ D {\bf 14} (1976) 3432
[Erratum-ibid.\ D {\bf 18} (1978) 2199].
%%CITATION = PHRVA,D14,3432;%%



\bibitem{Affleck} 
I.~Affleck,
``On Constrained Instantons,''
Nucl.\ Phys.\ B {\bf 191} (1981) 429.
%%CITATION = NUPHA,B191,429;%%


\bibitem{DKM-mo1} 
N.~Dorey, V.~V.~Khoze and M.~P.~Mattis,
``Multi-Instanton Calculus in N=2 Supersymmetric Gauge Theory,''
Phys.\ Rev.\ D {\bf 54} (1996) 2921
[arXiv:hep-th/9603136].
%%CITATION = HEP-TH 9603136;%%

\bibitem{DHKM-rept} 
N.~Dorey, T.~J.~Hollowood, V.~V.~Khoze and M.~P.~Mattis,
``The calculus of many instantons,''
Phys.\ Rept.\  {\bf 371} (2002) 231
[arXiv:hep-th/0206063].
%%CITATION = HEP-TH 0206063;%%


\bibitem{ADHM} 
M.~F.~Atiyah, N.~J.~Hitchin, V.~G.~Drinfeld and Y.~I.~Manin,
``Construction Of Instantons,''
Phys.\ Lett.\ A {\bf 65} (1978) 185.
%%CITATION = PHLTA,A65,185;%%



\bibitem{CGFT} 
E.~Corrigan, D.~B.~Fairlie, S.~Templeton and P.~Goddard,
``A Green's Function For The General Selfdual Gauge Field,''
Nucl.\ Phys.\ B {\bf 140} (1978) 31.
%%CITATION = NUPHA,B140,31;%%

  
\bibitem{CWS} 
N.~H.~Christ, E.~J.~Weinberg and N.~K.~Stanton,
``General Self-Dual Yang-Mills Solutions,''
Phys.\ Rev.\ D {\bf 18} (1978) 2013.
%%CITATION = PHRVA,D18,2013;%%


\bibitem{Y} 
A.~V.~Yung,
``Instanton Vacuum In Supersymmetric QCD,''
Nucl.\ Phys.\ B {\bf 297} (1988) 47.
%%CITATION = NUPHA,B297,47;%%

\bibitem{KR}
V.~V.~Khoze and A.~Ringwald,
``Nonperturbative contribution to total cross-sections in nonAbelian gauge theories,''
Phys.\ Lett.\ B {\bf 259} (1991) 106.
%%CITATION = PHLTA,B259,106;%%


\bibitem{JRCDG} 
R.~Jackiw and C.~Rebbi,
``Vacuum Periodicity In A Yang-Mills Quantum Theory,''
Phys.\ Rev.\ Lett.\  {\bf 37} (1976) 172;
%%CITATION = PRLTA,37,172;%%
\\
C.~G.~Callan, R.~F.~Dashen and D.~J.~Gross,
``The Structure Of The Gauge Theory Vacuum,''
Phys.\ Lett.\ B {\bf 63} (1976) 334.
%%CITATION = PHLTA,B63,334;%%


\bibitem{Manton} 
N.~S.~Manton,
``Topology In The Weinberg-Salam Theory,''
Phys.\ Rev.\ D {\bf 28} (1983) 2019.
%%CITATION = PHRVA,D28,2019;%%


\bibitem{Witten-i1}
E.~Witten,
``Sigma models and the ADHM construction of instantons,''
J.\ Geom.\ Phys.\  {\bf 15} (1995) 215
[arXiv:hep-th/9410052].
%%CITATION = HEP-TH 9410052;%%


\bibitem{Witten-i2}
E.~Witten,
``Small Instantons in String Theory,''
Nucl.\ Phys.\ B {\bf 460} (1996) 541
[arXiv:hep-th/9511030].
%%CITATION = HEP-TH 9511030;%%


\bibitem{Douglas-i1}
M.~R.~Douglas,
``Branes within branes,''
arXiv:hep-th/9512077.
%%CITATION = HEP-TH 9512077;%%

\bibitem{Douglas-i2}
M.~R.~Douglas,
``Gauge Fields and D-branes,''
J.\ Geom.\ Phys.\  {\bf 28} (1998) 255
[arXiv:hep-th/9604198].
%%CITATION = HEP-TH 9604198;%%


\bibitem{DHKMV-mo3} 
N.~Dorey, T.~J.~Hollowood, V.~V.~Khoze, M.~P.~Mattis and S.~Vandoren,
``Multi-instanton calculus and the AdS/CFT correspondence in N = 4  
superconformal field theory,''
Nucl.\ Phys.\ B {\bf 552} (1999) 88
[arXiv:hep-th/9901128].
%%CITATION = HEP-TH 9901128;%%


\bibitem{Diaconescu} 
D.~E.~Diaconescu,
``D-branes, monopoles and Nahm equations,''
Nucl.\ Phys.\ B {\bf 503} (1997) 220
[arXiv:hep-th/9608163].
%%CITATION = HEP-TH 9608163;%%

\bibitem{Hashimoto} 
A.~Hashimoto,
``The shape of branes pulled by strings,''
Phys.\ Rev.\ D {\bf 57} (1998) 6441
[arXiv:hep-th/9711097].
%%CITATION = HEP-TH 9711097;%%

\bibitem{Semenoff}
G.~W.~Semenoff and K.~Zarembo,
``Solitons on branes,''
Nucl.\ Phys.\ B {\bf 556}, 247 (1999)
[arXiv:hep-th/9903140].
%%CITATION = HEP-TH 9903140;%%

\bibitem{Harvey-lec} 
J.~A.~Harvey,
``Magnetic monopoles, duality, and supersymmetry,''
arXiv:hep-th/9603086.
%%CITATION = HEP-TH 9603086;%%

\bibitem{Osborn} 
H.~Osborn,
``Topological Charges For N=4 Supersymmetric Gauge Theories And Monopoles Of Spin 1,''
Phys.\ Lett.\ B {\bf 83} (1979) 321.
%%CITATION = PHLTA,B83,321;%%

\bibitem{MO} 
C.~Montonen and D.~I.~Olive,
``Magnetic Monopoles As Gauge Particles?,''
Phys.\ Lett.\ B {\bf 72} (1977) 117.
%%CITATION = PHLTA,B72,117;%%


\bibitem{FKS} 
E.~Farhi, V.~V.~Khoze and R.~J.~Singleton,
``Minkowski space nonAbelian classical solutions with noninteger winding number change,''
Phys.\ Rev.\ D {\bf 47} (1993) 5551
[arXiv:hep-ph/9212239].
%%CITATION = HEP-PH 9212239;%%

\bibitem{Harvey-sp} 
J.~A.~Harvey, P.~Horava and P.~Kraus,
``D-sphalerons and the topology of string configuration space,''
JHEP {\bf 0003} (2000) 021
[arXiv:hep-th/0001143].
%%CITATION = HEP-TH 0001143;%%


\bibitem{instquarks} 
 A.A. Belavin, V.A. Fateeev, A.S. Schwarz and Y.S. Tyupkin, 
Phys. Lett. {\bf 83B} (1979) 317;
\\
 V.A. Fateeev, I.V. Frolov and A.S. Schwarz, 
Nucl. Phys. {\bf B154} (1979) 1;
\\
B. Berg and M. L{\"u}scher, 
Comm. Math. Phys. {69} (1979) 57;
\\
H. Osborn, Ann. Phys. {\bf 135} (1981) 373.


\bibitem{DHKM-mono} 
N.~M.~Davies, T.~J.~Hollowood, V.~V.~Khoze and M.~P.~Mattis,
``Gluino condensate and magnetic monopoles in supersymmetric  gluodynamics,''
Nucl.\ Phys.\ B {\bf 559} (1999) 123
[arXiv:hep-th/9905015].
%%CITATION = HEP-TH 9905015;%%


\bibitem{DHK-mono} 
N.~M.~Davies, T.~J.~Hollowood and V.~V.~Khoze,
``Monopoles, affine algebras and the gluino condensate,''
J.\ Math.\ Phys.\  {\bf 44} (2003) 3640
[arXiv:hep-th/0006011].
%%CITATION = HEP-TH 0006011;%%


\bibitem{Yi-cyl} 
K.~M.~Lee and P.~Yi,
``Monopoles and instantons on partially compactified D-branes,''
Phys.\ Rev.\ D {\bf 56}, 3711 (1997)
[arXiv:hep-th/9702107].
%%CITATION = HEP-TH 9702107;%%


\bibitem{Nahmtwo}
W. Nahm, ``Self-dual Monopoles and Calorons'',
in: Lecture Notes in Physics 201, Eds. G. Denado {\it et. al.} 1984,
p. 189.

\bibitem{KvBaal} 
T.~C.~Kraan and P.~van Baal,
``Exact T-duality between calorons and Taub - NUT spaces,''
Phys.\ Lett.\ B {\bf 428} (1998) 268
[arXiv:hep-th/9802049];
%%CITATION = HEP-TH 9802049;%%
\\
``Periodic instantons with non-trivial holonomy,''
Nucl.\ Phys.\ B {\bf 533} (1998) 627
[arXiv:hep-th/9805168].
%%CITATION = HEP-TH 9805168;%%

\bibitem{S-branes} 
M.~Gutperle and A.~Strominger,
``Spacelike branes,''
JHEP {\bf 0204} (2002) 018
[arXiv:hep-th/0202210].
%%CITATION = HEP-TH 0202210;%%


\bibitem{125} 
C.~M.~Hull and P.~K.~Townsend,
``Unity of superstring dualities,''
Nucl.\ Phys.\ B {\bf 438} (1995) 109
[arXiv:hep-th/9410167].
%%CITATION = HEP-TH 9410167;%%

\bibitem{127}
 E.~Witten,
``String theory dynamics in various dimensions,''
Nucl.\ Phys.\ B {\bf 443} (1995) 85
[arXiv:hep-th/9503124].
%%CITATION = HEP-TH 9503124;%%

\bibitem{110} 
J.~H.~Schwarz,
``An SL(2,Z) multiplet of type IIB superstrings,''
Phys.\ Lett.\ B {\bf 360} (1995) 13
[Erratum-ibid.\ B {\bf 364} (1995) 252]
[arXiv:hep-th/9508143].
%%CITATION = HEP-TH 9508143;%%

\bibitem{PolchinskiII} 
J.~Polchinski,
String Theory. Vol. 2,
 Cambridge University Press 1998.

\bibitem{CVJ} 
C.~V.~Johnson,
``D-brane primer,''
arXiv:hep-th/0007170.
%%CITATION = HEP-TH 0007170;%%


\bibitem{Dinst} 
G.~W.~Gibbons, M.~B.~Green and M.~J.~Perry,
``Instantons and Seven-Branes in Type IIB Superstring Theory,''
Phys.\ Lett.\ B {\bf 370} (1996) 37
[arXiv:hep-th/9511080];
%%CITATION = HEP-TH 9511080;%%
\\
M.~B.~Green,
``Interconnections between type II superstrings, M theory and N = 4  Yang-Mills,''
arXiv:hep-th/9903124.
%%CITATION = HEP-TH 9903124;%%

\bibitem{47} 
C.~G.~Callan and J.~M.~Maldacena,
``Brane dynamics from the Born-Infeld action,''
Nucl.\ Phys.\ B {\bf 513} (1998) 198
[arXiv:hep-th/9708147].
%%CITATION = HEP-TH 9708147;%%

\bibitem{46} 
G.~W.~Gibbons,
``Born-Infeld particles and Dirichlet p-branes,''
Nucl.\ Phys.\ B {\bf 514} (1998) 603
[arXiv:hep-th/9709027].
%%CITATION = HEP-TH 9709027;%%

\bibitem{48} 
P.~S.~Howe, N.~D.~Lambert and P.~C.~West,
``The self-dual string soliton,''
Nucl.\ Phys.\ B {\bf 515} (1998) 203
[arXiv:hep-th/9709014].
%%CITATION = HEP-TH 9709014;%%

 
\bibitem{Gross-sp} 
N.~Drukker, D.~J.~Gross and N.~Itzhaki,
``Sphalerons, merons and unstable branes in AdS,''
Phys.\ Rev.\ D {\bf 62} (2000) 086007
[arXiv:hep-th/0004131].
%%CITATION = HEP-TH 0004131;%%

  
\end{thebibliography}
\end{document}